\documentclass[%
 reprint,
superscriptaddress,
nofootinbib,
bibnotes,
 amsmath,amssymb,
 aps,
]{revtex4-1}

\usepackage{latexsym,mathrsfs,bbm}
\usepackage{tikz}
\usetikzlibrary{backgrounds,fit,decorations.pathreplacing,calc}
\usepackage{graphicx}
\usepackage{dcolumn}
\usepackage{braket}
\usepackage{gensymb}
\usepackage{bm}
\usepackage{amsmath}
\usepackage{amsthm}
\usepackage{float}
\usepackage{cancel}
\usepackage{mathtools}
\usepackage{color}
\usepackage{csquotes}
\usepackage{framed} 
\usepackage[utf8]{inputenc}
\usepackage{thmtools}

\usepackage{diagbox}
\usepackage{adjustbox}

\theoremstyle{plain}

\newcommand{\bT}{{\theta}}

\newcommand{\pps}{p_{\bT}^{\mathrm{PS}}}

\newcommand{\LParen}{ \bm{(} }
\newcommand{\RParen}{ \bm{)} }

\usepackage{hyperref}

\allowdisplaybreaks

\begin{document}

\preprint{APS/123-QED}

\title{Nonclassical advantage in metrology established via quantum simulations of
hypothetical closed timelike curves}

\author{David R. M. Arvidsson-Shukur}
\affiliation{
Hitachi Cambridge Laboratory, J. J. Thomson Avenue,  Cambridge,  CB3 0HE,  United Kingdom
}

\author{Aidan G.  McConnell}
\affiliation{%
 Cavendish Laboratory, Department of Physics, University of Cambridge, Cambridge, CB3 0HE, United Kingdom
}%
\affiliation{ Laboratory for X-ray Nanoscience and Technologies, Paul Scherrer Institut, 5232 Villigen, Switzerland}
\affiliation{ Department of Physics and Quantum Center, Eidgenössische Technische Hochschule Zürich, CH-8093 Z\"urich, Switzerland }

\author{Nicole Yunger Halpern}
\affiliation{Joint Center for Quantum Information and Computer Science, NIST and University of Maryland, College Park, MD 20742, USA}
\affiliation{Institute for Physical Science and Technology, University of Maryland, College Park, MD 20742, USA}

\date{\today}

\begin{abstract}
We construct a metrology experiment in which the metrologist can sometimes amend her input state by simulating a closed timelike curve, a worldline that travels backward in time. The existence of closed timelike curves is hypothetical. Nevertheless, they can be simulated probabilistically by quantum-teleportation circuits. We leverage such simulations to pinpoint a counterintuitive nonclassical advantage achievable with entanglement. Our experiment echoes a common information-processing task: A metrologist must prepare probes to input into an unknown quantum interaction. The goal is to infer as much information per probe as possible. If the input is optimal, the information gained per probe can exceed any value achievable classically. The problem is that, only after the interaction does the metrologist learn which input would have been optimal. The metrologist can attempt to change her input by effectively teleporting the optimal input back in time, via entanglement manipulation. The effective time travel sometimes fails but ensures that, summed over trials, the metrologist’s winnings are positive. Our Gedankenexperiment demonstrates that entanglement can generate operational advantages forbidden in classical chronology-respecting theories. 
\end{abstract}

\maketitle


\textit{Introduction}.---The arrow of time makes  gamblers, investors,  and quantum experimentalists perform actions that, in hindsight, are suboptimal.  Examples arise in  quantum metrology, the field of using nonclassical phenomena to estimate unknown parameters \cite{Giovanetti11}. The optimal input states and final measurements are often  known only once the experiment has finished \cite{Salmon22}.  Below, we present a Gedankenexperiment that circumvents this problem via  entanglement manipulation.

\begin{figure}
\includegraphics[scale=0.19]{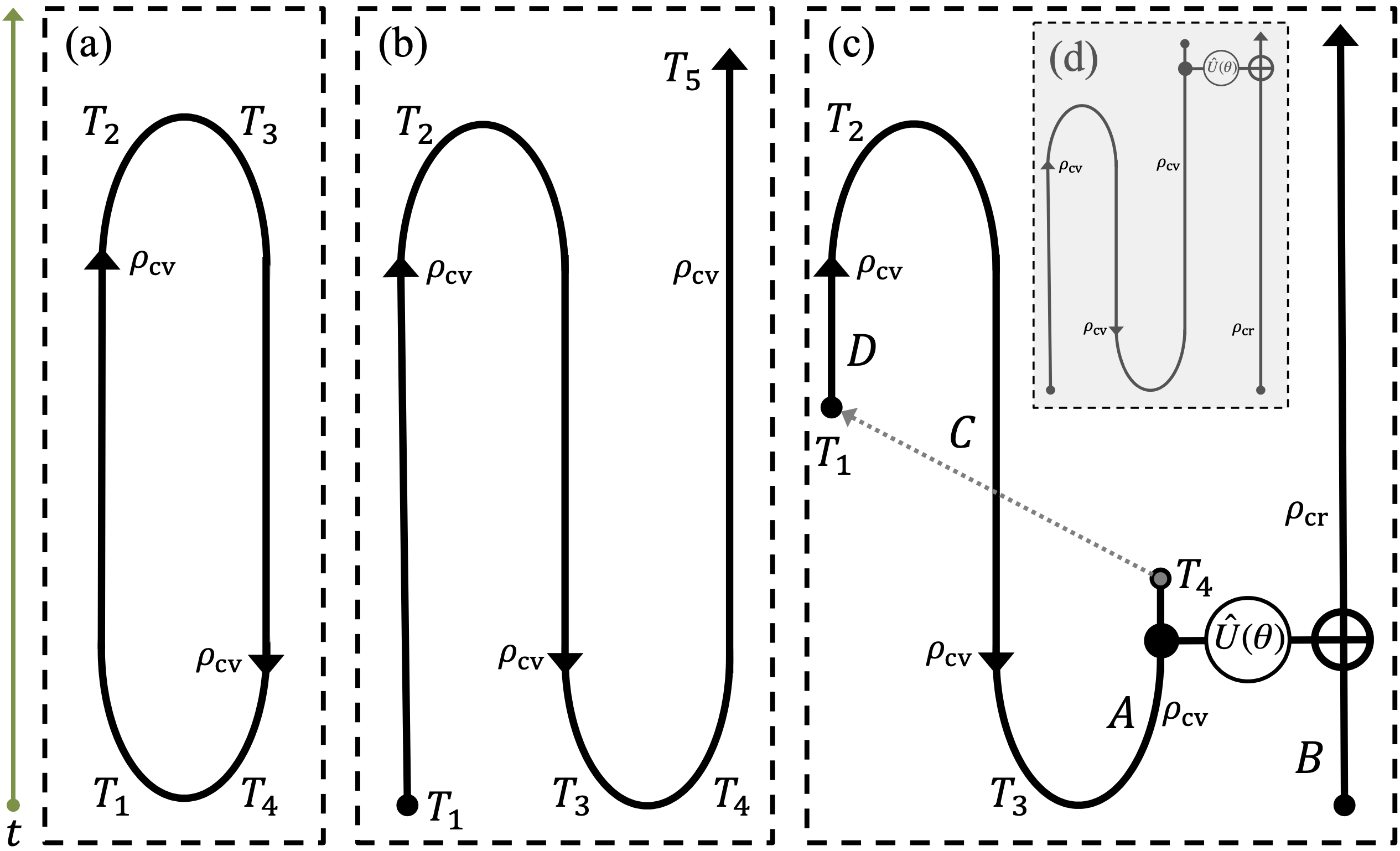}
\caption{Examples of chronology-violating particles traversing  hypothetical CTCs.   $\rho_{\mathrm{cv}}$ denotes such particles' states.  Time $t$, experienced by a chronology-respecting observer, runs from bottom to top. The time-traveling particle experiences time $T$.  (a) Closed loop. (b) $\rho_{\mathrm{cv}}$ returns to its past and then travels forward in time again. (c) CTC interpretation of the successful trials of our Gedankenexperiment.  $\rho_{\mathrm{cv}}$  is created at $T_1$ and travels forward in time until $T_2$. Then,  it reverses  temporal direction and travels backward in time until reaching $T_3$. After that,  it again travels forward in time.  $\rho_{\mathrm{cv}}$ then interacts with a chronology-respecting state, $\rho_{\mathrm{cr}}$,  and is subsequently destroyed, prior to $\rho_{\mathrm{cv}}$'s creation ($T_1$).  For comparison, the inset (d) depicts the standard teleportation, across space, of a quantum state needed as input for an interaction.}
\label{fig:CTCExamples}
\end{figure}

A common metrological goal is to estimate the strength of a weak interaction between a system in a state $\ket{\phi}$ and a probe in a state $\ket{\psi}$.  The interaction strength can be estimated from the data from several measured probes.   Upon measuring probes at too high an intensity, detectors can saturate---cease to function until given time to reset \cite{Jordan14, Dressel14, Harris17, ArvShukur19-2,Lupu22,Semenov23}.  Additionally, one might lack the memory needed to store all the probes \cite{Jenne21} , or lack the computational power needed to process the probes' contents after their measurement \cite{Scandi23}.   Reducing the number of probes measured is therefore often advantageous \cite{Dressel14, Pang14,  Harris17,  Xu20,  ArvShukur19-2, ArvShukur21, Lupu22, Salvati23}.    In such situations, one can use \textit{weak-value amplification} to boost the amount of information obtained per measured probe \cite{Vaidman88, Duck89, Hosten08, Dressel14, Pang14,  Harris17,  Xu20}. In weak-value amplification,    the system is initialized in a state $\ket{\phi_{\mathrm{i}}}$, the system interacts with the probe, and then the system is measured.  If, and only if, the  system's measurement outcome corresponds to $\ket{\phi_{\mathrm{f}}}$, the probe is measured.  Successful pre- and postselection guarantees that the probe carries a large amount of information.  Weak-value amplification  stems from genuine nonclassicality, as reviewed below \cite{Tollaksen07, Pusey14, Kunjwal19}.  The nonclassicality originates in both the postinteraction measurement and the initialization, sparking discussions about chronology-violating physics \cite{Aharonov08, Leifer17}.

Chronology-violating physics includes  \textit{closed timelike curves} (CTCs) \cite{Godel49, Morris88, Deutsch91, Bennett05, Svetlichny11, Lloyd11, Lloyd11-2}. A CTC is a hypothetical spacetime worldline that loops backward in time (Fig. \ref{fig:CTCExamples}). Particles   that follow CTCs  could travel backward in time with respect to chronology-respecting observers.  Although allowed by general relativity, CTCs lead to logical paradoxes. A famous example is the \textit{grandfather paradox}: A time traveler travels back in time to kill her grandfather, before he fathers any children, such that the time traveler could never have been born$\ldots$ Such inconsistency can characterize  classical CTCs and has prompted scientific discussions about CTCs' likelihood of existing \cite{Gott91, Deutsch91, Hawking92, Deser92,  Carroll94, Lloyd11-2}. Two competing theories resolve  such paradoxes, self-consistently reconciling general-relativistic CTCs with quantum theory \cite{Deutsch91, Bennett05, Svetlichny11, Lloyd11, Lloyd11-2, Allen14,  Brun17}.  We use the theory of \emph{postselected CTCs} (PCTCs). PCTCs are equivalent to quantum circuits that involve \emph{postselection}, or conditioning on certain measurement outcomes \cite{Lloyd11-2}. Such circuits have been realized experimentally \cite{Lloyd11}. Our results concern postselected circuits that achieve weak-value amplification.



In this work,  we show that postselected quantum-teleportation circuits can  effectively send useful states from the future to the past,  providing access to nonclassical phenomena in quantum metrology. We propose a  weak-value-amplification Gedankenexperiment for estimating the strength of an interaction between a system and a probe.  Motivated by the aforementioned practical limitations, the figure of merit is the amount of information obtained per probe. 
As mentioned earlier, this rate can be  nonclassically large if one discards the probe conditionally on an earlier measurement of the system.
But this  information distillation requires that the systems be  initialized in a specific state. In our Gedankenexperiment, the  optimal input state is unknown until after the system has been measured.   We circumvent this challenge  via  quantum theory's ability to simulate backward time travel: One can effectively teleport the optimal state from the experiment's end to its beginning.  The simulated time travel sometimes fails, but at no detriment to the figure of merit, the amount of information gleaned from the remaining probes.  These probes, retained only if the simulated time travel succeeds, carry amounts of information impossible to achieve classically.  Thus, in weak-value amplification, the system can be initialized  \textit{after} the  system--probe interaction---paradoxically, in chronology-respecting theories.  Our conceptual results  pinpoint a deep connection between  entanglement and   effectively retrocausal correlations that enable nonclassical advantages.

\textit{Background: Closed timelike curves}.---Figure \ref{fig:CTCExamples} shows examples of CTCs---hypothetical spacetime worldlines that loop in the direction of time.  Two (primary) theories entail self-consistent quantum descriptions of CTCs. The first theory is called Deutsch's CTCs (DCTCs) \cite{Deutsch91, Svetlichny11}.   DCTCs conserve a time traveler's state but not the state's correlations (e.g., entanglement) with chronology-respecting systems. 


We use a second model: PCTCs \cite{Bennett05, Svetlichny11, Lloyd11, Lloyd11-2, Allen14,  Brun17},  which cast CTCs as quantum communication channels to the past \cite{ Lloyd11-2}. The following condition defines PCTCs: Consider measuring a system that undergoes a PCTC.  Whether the measurement happens before or after the PCTC does not affect the measurement statistics.  Such self-consistency follows from modeling CTCs with  quantum-teleportation circuits (quantum-communication channels) that involve postselection. The postselection ensures that time-traveling particles preserve their correlations with chronology-respecting systems.

Quantum circuits with entangled inputs can effectively realize PCTCs, as illustrated in Fig. \ref{fig:CTCExamples}.  (The word ``effectively'' is used because  one cannot empirically prove whether time travel \textit{actually} happened \cite{Svetlichny11}.)   There,  the   $\cup$ depicts the creation of a Bell (maximally entangled) state \cite{Nielsen11}. The  $\cap$ depicts the future postselection on that Bell state.  In  Fig. \ref{fig:CTCExamples}(a), the two entangled particles  can  be viewed as the forward-traveling (left) and backward-traveling (right) parts of one chronology-violating particle's worldline. 

The CTC in Fig. \ref{fig:CTCExamples}(b) can be simulated by a three-qubit quantum-teleportation circuit.    With probability $1/4$,  the to-be teleported qubit appears at the receiver's end, without the receiver's performing any local operation  \cite{Bennett93}. In these events, the teleported qubit was already at the receiver's end \cite{Bennett05,  Coecke04, Lloyd11}.\footnote{A teleportation scheme that takes all possible outcomes into account,  without the receiver's performing any local operations, transports an unknown qubit exactly as ineffectively as random guessing. } 
 Postselected on these outcomes, the circuit can be viewed as mimicking  one chronology-violating qubit's worldline. In the  qubit's rest frame, the qubit is  initialized at $T_1$.  At $T_2$,  it starts traveling backward according to the laboratory frame, until reaching the point of its ``birth'' at $T_3$.  At $T_4$, the qubit reverses its temporal direction again, returning to traveling forward in time. 

We do not argue for or against the physical existence of PCTCs.  Rather, we identify a consequence of quantum theory's ability to simulate PCTCs: a counterintuitive metrological advantage achievable with entanglement. Below, we outline a Gedankenexperiment  that achieves this advantage.  First, we review weak-value amplification in quantum metrology.

\textit{Weak-value amplification for metrology}.---We now describe how to estimate the strength of a weak system--probe interaction.  Weak-value amplification concentrates information, boosting  the amount of information obtained per probe. 

Using quantum metrology, one infers the value of an unknown parameter $\theta$ by measuring $N$ copies of a  state  $\ket{\Psi_\theta}$ \cite{Giovanetti11}.  Every such procedure implies  an estimator $\theta_{\mathrm{e}}$ of $\theta$.  The Cramér-Rao inequality lower-bounds the precision of every unbiased  $\theta_{\mathrm{e}}$:
\begin{equation}
\label{Eq:CR}
\mathrm{Var}(\theta_{\mathrm{e}}) \geq \frac{1}{N \cdot \mathcal{I}_{\mathrm{q}}(\theta | \Psi_\theta )} .
\end{equation}
$\mathcal{I}_{\mathrm{q}}(\theta | \Psi_\theta )$ is the \textit{quantum Fisher information}, which quantifies the average amount of information learned about  $\theta$  per optimal measurement \cite{Braunstein94}. The quantum Fisher information has the form 
\begin{equation}
\label{Eq:FisherPure}
\mathcal{I}_{\mathrm{q}}(\theta | \Psi_\theta ) = 4\braket{ \dot{\Psi}_\theta |  \dot{\Psi}_\theta} - 4 | \braket{ \Psi_\theta | \dot{\Psi}_\theta} |^2 ,
\end{equation}
where $\dot{x} \equiv \partial x / \partial \theta$.  Common estimators saturate Ineq. \eqref{Eq:CR}  when $N$ is large.  The larger $\mathcal{I}_{\mathrm{q}}(\theta | \Psi_\theta )$ is, the more precisely one can estimate $\theta$. 

In this work, we consider  estimating the strength of an interaction $\hat{U}(\theta) = e^{- i \theta \hat{\Pi}_{a} \otimes \hat{B} / 2}$  between a system qubit in a state $\ket{\phi}_{A} $ and a probe qubit in a state $\ket{\psi}_{B}$.\footnote{We choose this simple, common two-qubit interaction to align with previous works \cite{Pusey14, Kunjwal19} and to illustrate our metrological advantage.  More-general interactions imply analogous results, at the expense of more-tedious analytics.} Here, $\theta \approx 0$ is the weak-coupling strength, and $\hat{\Pi}_{a} \equiv \ket{a}\bra{a}$ denotes a rank-$1$ projector on qubit $A$'s  Hilbert space.  $\hat{B} \equiv  \ket{b^+}\bra{b^+}  -  \ket{b^-}\bra{b^-}  $ is a Hermitian operator acting on qubit $B$'s Hilbert space, with eigenvalues $\pm1$. $\hat{U}(\theta)$  evolves $\ket{\psi}_{B}$ with a unitary evolution generated by $\hat{B}$, conditionally on qubit $A$'s being in the  state $\ket{a}$.  

To measure the coupling strength $\theta$, we  prepare the system-and-probe state $\ket{\Psi_0 }_{A,B}  \equiv \ket{\phi}_{A} \ket{\psi}_{B}$, evolve it under $\hat{U}(\theta)$, and then measure the qubits.  An information-optimal input is $ \ket{\Psi_0^{\star} }_{A,B}  = \ket{a}_{A} \frac{1}{\sqrt{2} } ( \ket{b^+}_{B} + \ket{b^-}_{B} )$; this state acquires the greatest  possible quantum Fisher information, being maximally sensitive  to changes in $\theta$.  The postinteraction state is 
\begin{align}
\label{Eq:PrePSState}
 \ket{\Psi^{\star}(\theta)}_{{A,B}} & \equiv  \hat{U}(\theta) \ket{\Psi_0^{\star} }_{{A,B}} \nonumber \\ & = \ket{a}_{{A}}  \frac{ e^{ - i \theta /2}  \ket{b^+}_{{B}}  + e^{ i \theta /2}  \ket{b^-}_{{B}} }{\sqrt{2}} .
\end{align}
According to Eq. \eqref{Eq:FisherPure},  the average measurement yields $\mathcal{I}_{\mathrm{q}}[\theta | \Psi^{\star}_{{A,B}} (\theta) ] = 1$ unit of Fisher information per postinteraction state.

\begin{figure}
\includegraphics[scale=0.23]{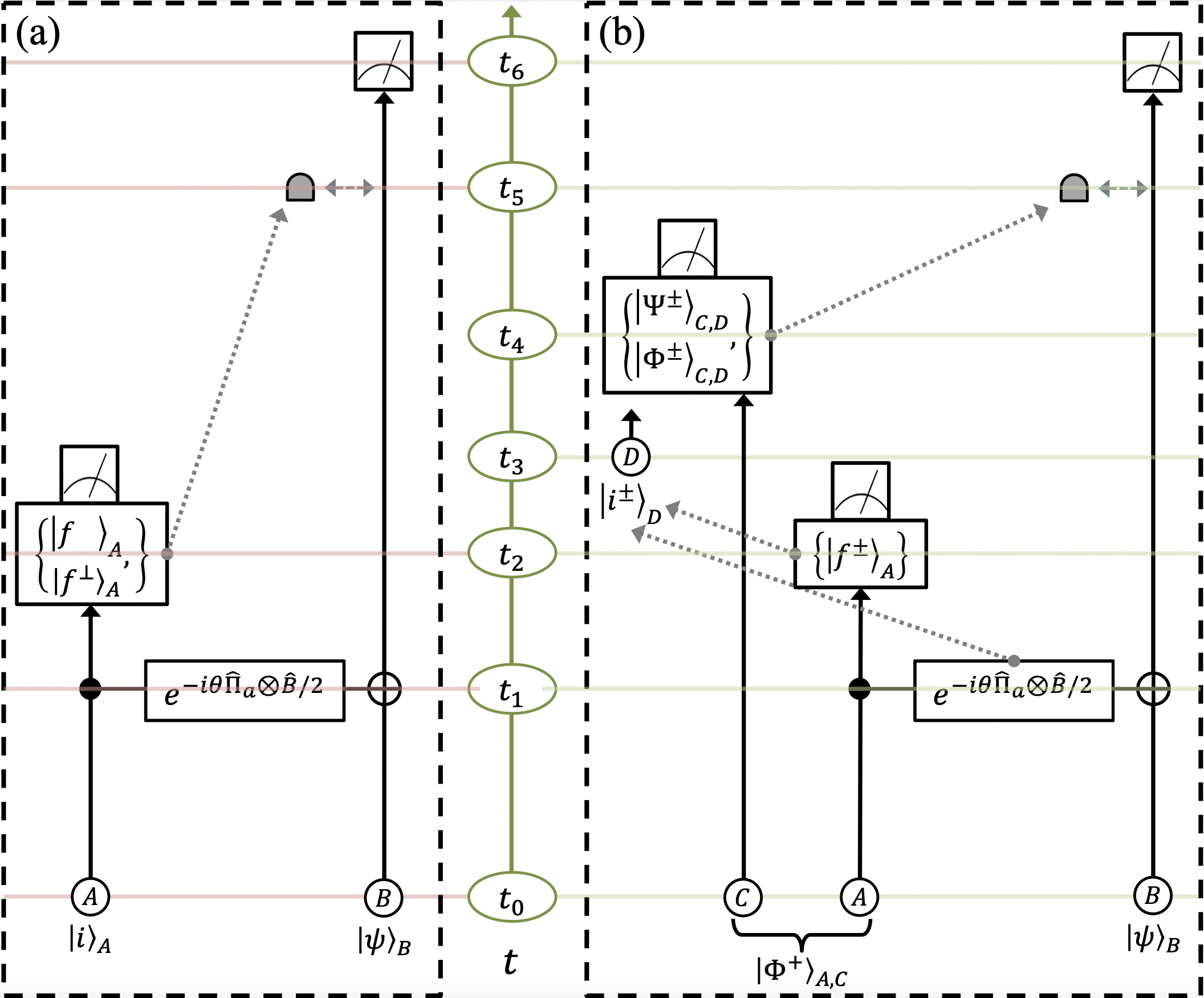}
\caption{Circuit diagrams for (a) standard and (b) PCTC-powered weak-value amplification. Time progresses in the laboratory's rest frame as one proceeds upward along the central, vertical axis.  Black lines represent qubits. Dashed gray lines represent classical information.  }
\label{fig:CircuitRep}
\end{figure}

Usefully, one can distill much information into few probes.  One measures system $A$ and, conditionally on the outcome, discards  or keeps (postselects) the probe $B$.  Information distillation is particularly advantageous if one's detectors  saturate, if memory constraints limit one's data storage, or if computational resources limit one's postprocessing power.   Then, qubit $B$ merits measuring only if $B$  carries much information  \cite{Jordan14, Dressel14, Harris17, ArvShukur19-2,Lupu22,Semenov23}.  We now review one such distillation scheme, \textit{weak-value amplification}  \cite{Dressel14, Pang14,  Harris17,  Xu20, Vaidman88, Duck89, Hosten08},  depicted in Fig.  \ref{fig:CircuitRep}(a).

One evolves $\ket{\Psi_0^{\mathrm{w}} }_{{A,B}}  \equiv \ket{i}_{{A}} \frac{1}{\sqrt{2} } ( \ket{b^+}_{{B}} + \ket{b^-}_{{B}} )$ under $\hat{U}(\theta)$,  then  measures $A$ in the basis $\{ \ket{f}, \ket{f^{\perp}}\}$.  If the outcome is $_{{A}}\bra{f}$,  the blocker in Fig. \ref{fig:CircuitRep}(a) is removed,  and  $B$ is measured. If not,  the blocker destroys $B$  The postselected state is 
\begin{align}
\label{Eq:PSstate}
\ket{\Psi^{\rm{PS}} (\theta)}_{{B}}=  \ket{\psi^{\rm{PS}} (\theta)}_{{B}} / \sqrt{p_{\theta}^{\mathrm{PS}} },
\end{align}
where $\ket{\psi^{\rm{PS}}  (\theta)}_{{B}}\equiv \left( _{{A}}\bra{f} \otimes \hat{1}_{{B}} \right) \hat{U}(\theta) \ket{\Psi_0^{\mathrm{w}} }_{{A,B}} $. The probability of postselecting successfully is $p_{\theta}^{\mathrm{PS}}  \equiv {}_{B}\braket{\psi^{\rm{PS}}  (\theta)|\psi^{\rm{PS}}  (\theta)}_{{B}}$.  A little algebra simplifies the postselected state,  if $|\theta \cdot  _{f}\braket{\hat{\Pi}_{a}}_{i} | \ll 1$:
\begin{align}
\ket{\Psi^{\rm{PS}} (\theta)}_{{B}}=  & \frac{ e^{ - i \theta _{f}\braket{\hat{\Pi}_{a}}_{i} /2  }  \ket{b^+}_{{B}} +   e^{ i \theta _{f}\braket{\hat{\Pi}_{a}}_{i} /2  } \ket{b^-}_{{B}} }{\sqrt{2} }   \nonumber \\ & + \mathcal{O}(\theta^2) .
\end{align}
  The  \textit{weak value} of $\hat{\Pi}_{a} $ is 
\begin{equation}
\label{Eq:WV}
_{f}\braket{\hat{\Pi}_a}_{i} \equiv \frac{_{{A}}\bra{f} \hat{\Pi}_{a} \ket{i}_{{A}}  }{ _{{A}}\braket {f | i}_{{A}} } ,
\end{equation}
 the  ``expectation value'' of $\hat{\Pi}_{a}$ preselected on the state $\ket{i}_{{A}} $ and postselected on  $_{{A}}\bra{f}$.  The quantum Fisher information [Eq. \eqref{Eq:FisherPure}] of $\ket{\Psi^{\rm{PS}} (\theta)}_{B}$ is 
\begin{align}
\label{Eq:WeakPSFish}
 \mathcal{I}_{\mathrm{q}} \LParen \theta | \Psi^{\rm{PS}}_{{B}} (\theta) \RParen =
 \left| _f\braket{\hat{\Pi}_{a}}_{i} \right|^2 + \mathcal{O}\left( \theta \right)
   .
\end{align}

Above, we found that the nonpostselected experiment's quantum Fisher information,  $\mathcal{I}_{\mathrm{q}} \LParen \theta |\Psi_{{A,B}} (\theta) \RParen$, has a maximum value of $1$. The postselected experiment, however, can achieve a quantum Fisher information  $\mathcal{I}_{\mathrm{q}} \LParen \theta | \Psi^{\rm{PS}}_{{B}} (\theta) \RParen \gg 1$.  Weak-value amplification does not increase the total amount of information gained from all the probes \cite{Combes14, Ferrie14-2},  but distills large amounts of information into a few  postselected probes. 

Such anomalously large amounts of information  witness nonclassical phenomena \cite{ArvShukur19-2, ArvShukur21, ArvShuk21-2, Lupu22}. For small $\theta$, Eq. \eqref{Eq:WeakPSFish} excedes $1$ if, and only if, the weak value $| _f\braket{\hat{\Pi}_{a}}_{i}| > 1$, i.e., the weak value's magnitude exceeds the greatest eigenvalue of $\hat{\Pi}_{a}$.   Such a weak value is called \emph{anomalous}.  Anomalous weak values arise from the quantum resource \emph{contextuality}: One can try to model quantum systems as being in real, but unknown, microstates like microstates in classical statistical mechanics. In such a framework, however, operationally indistinguishable quantum procedures cannot be modeled identically. This impossibility is contextuality \cite{Spekkens05, Pusey14,  Kunjwal19, Lostaglio20},  which  is valuable.  It enables weak-value amplification,  which  compresses many probes' metrological information into a few highly informative probes.


\textit{Metrological quantum advantage via PCTC simulation}.---To perform weak-value amplification, an experimentalist  must carefully  choose qubit $A$'s input state, $\ket{i}_{{A}}$, and  final-measurement basis, $\{ \ket{f}_{{A}},  \ket{f^{\perp}}_{{A}}\}$, such that $| _f\braket{\hat{\Pi}_{a}}_{i} | > 1$.  Doing so requires knowledge of $\hat{\Pi}_{a}$.  The goal is to simultaneously achieve a small weak-value denominator $_{{A}}\braket {f | i}_{{A}}$ and large numerator $_{{A}}\bra{f} \hat{\Pi}_{a} \ket{i}_{{A}} $  in Eq. \eqref{Eq:WV}.  If $\hat{\Pi}_{a}$ and the postselection basis are unknown,  achieving the goal seems impossible.

We overcome this obstacle by  implementing postselected metrology with a  quantum circuit that simulates a PCTC.  We  assume that $ \hat{\Pi}_{a}$ and the postselection basis $\{ \ket{f^{\pm}}_{{A}} \}$ are unknown until just after (in the laboratory's rest frame) the interaction.\footnote{Even if our delay's precise length is chosen artificially, our story cleaves to the real-world principle that metrology tends to involve delays.  For example,  optimal input states are generally known only once the specifics of the interaction are known \cite{Braunstein94,Giovanetti06, Salmon22}.}  Can we nevertheless initialize $\ket{\phi}_{{A}}$ to leverage   contextuality? We answer affirmatively, by constructing a  PCTC simulation.

Given a  postelection  outcome $_{{A}} \bra{f^{\pm}}$, we choose the  input state  $\ket{i^{\pm}}_{{A}}$ such that $| _{f^+}\braket{\hat{\Pi}_{a}}_{i^+} | = | _{f^-}\braket{\hat{\Pi}_{a}}_{i^-} |  \gg 1$.\footnote{In general, $\braket{i^- | i^+} \neq 0$.} As $ \hat{\Pi}_{a}$ and  $\{ \ket{f^{\pm}}_{{A}} \}$ are known only after the interaction, we effectively create the  input state $\ket{\phi}_{{A}} = \ket{i^{\pm}}_{{A}}$ after the interaction has taken place. Then,   using a postselected quantum-teleportation circuit, we  effectively transport the state backward in time, such that $\ket{\phi}_{{A}}$  serves as an input  to the interaction.  Figure \ref{fig:CircuitRep}(b) illustrates our experiment with a quantum circuit. 


In the laboratory's rest frame, our protocol proceeds as follows:
\begin{enumerate}

\item[$t_0$: ]{
$\bullet$
$A$ and $C$ are entangled: $$\ket{\Phi^{+}}_{{A},{C}}= \frac{1}{\sqrt{2}} \left( \ket{0}_{{A}} \ket{0}_{{C}} +  \ket{1}_{{A}}\ket{1}_{{C}}\right).$$

$\bullet$ Qubit $B$ is initialized to $$\ket{\Psi}_{{B}} =  \frac{1}{\sqrt{2} } \left( \ket{b^+}_{{B}} + \ket{b^-}_{{B}} \right) .$$
}

\item[$t_1$: ]{$\bullet$ $A$ and $B$  interact via $\hat{U}(\theta)=e^{- i \theta \hat{\Pi}_{a} \otimes \hat{B} / 2}$. The value of $\theta$ and the form of $\hat{\Pi}_{a} = \ket{a}\bra{a}$ are unknown.  }

\item[$t_2$: ]{
$\bullet$ The as-yet-unknown, optimal measurement basis $\{ \ket{f^{\pm}}_{{A}} \}$ is revealed. 

$\bullet$ $A$ is measured in this basis. 
}

\item[$t_3$: ]{$\bullet$ Information about $\hat{\Pi}_{a}$ and about the outcome $_{{A}} \bra{f^{\pm}}$  reaches $D$. 

$\bullet$ Qubit $D$ is created and initialized in $\ket{i^{\pm}}_{{D}}$.}

\item[$t_4$: ]{$\bullet$ $C$ and $D$ are measured in the Bell basis \cite{Nielsen11}. 

$\bullet$ Outcome $_{{C},{D}} \bra{\Phi^+}$ effectively teleports $\ket{i^{\pm}}_{{D}}$ to the time-$t_0$ system $A$.\footnote{This postselection is onto a copy of the state prepared initially, $\ket{\Phi^+}$. Therefore, our simulated CTC preserves correlations (between chronology-violating and chronology-respecting systems), as required by the definition of PCTCs.}}

\item[$t_5$: ]{
$\bullet$ If, and only if, outcome $_{{C},{D}} \bra{\Phi^+}$ was obtained at $t_4$,  a beam blocker is removed from $B$'s path.  }

\item[$t_6$: ]{$\bullet$ If the beam blocker was removed, $B$ is measured in the $\left\{ \frac{1}{\sqrt{2}} ( \ket{b^+} \pm \ket{b^-} )  \right\}$ basis.}

\end{enumerate}
\noindent Supplementary Note I presents the  mathematical details behind our protocol's effectiveness.  Supplementary Note II (which  references experimental works \cite{Barbieri05, Cinelli05, Barreiro05, Vallone07, Chen07, Ciampini16}) proposes an optics realization.


Repeated experiments that involve final $B$ measurements  produce an anomalously large weak value.  Hence, our scheme amplifies the quantum Fisher information about $\theta$ to nonclassically large values.  We have thus shown that, in weak-value amplification, the preselected system state can effectively be created \textit{after} the  interaction---even after the state has been measured and destroyed.  This point is visible in Fig.  \ref{fig:CTCExamples}(c),  a CTC depiction of our protocol. The inset [Fig. \ref{fig:CTCExamples}(d)]  shows the standard teleportation, across space, of a quantum state to be inputted into an interaction. The state's initialization is postponed, and the state's destruction is advanced, in Fig.   \ref{fig:CTCExamples}(c).  These changes do not affect the chronology-respecting state's final form.  In each of many previous studies, classical or quantum information---but not both---traverses a PCTC.  Our study differs.  Quantum information (solid line) and  classical information (dashed line) form the loop of our  simulated CTC.

One could imagine three objections. First, some postselections---and so teleportation attempts---fail. However, these failures do not lower the figure of merit,   the average amount of information per probe that passes the blocker.  As further reassurance: Our setup does not send classical information  to the past.  [The dashed line in Fig.   \ref{fig:CTCExamples}(c) travels only forward in time.] The improved information-per-detection rate is available only at the end of the experiment.

Second, one might view as artificial our assumption about when the information needed to choose $\ket{i^{\pm}}$ arrives.  Indeed, the assumption is artificial.  Our study's purpose is foundational---to demonstrate the power of entanglement to achieve a counterintuitive metrological advantage. Nevertheless, our setup illustrates the general metrological principle that optimal input states  are known only once  the specifics of the interaction $\hat{U}(\theta)$ are known \cite{Braunstein94,Giovanetti06}.
However, Supplementary Notes III and IV contain two extensions of our protocol: one extension with a greater success probability and one extension with greater practicality.


Third, one might view our experiment as involving a preselected state $\ket{\Phi^{+}}_{{C},{A}} \ket{i^{\pm}}_{{D}}$ and a postselected state $_{{A}} \bra{f^{\pm}} _{{D},{C}} \bra{\Phi^+}$.  Our experiment would entail no more effective retrocausality than earlier experiments with pre- and postselection. However, such an interpretation contradicts the definitions of pre- and postselection, as $D$ is created after $A$ is postselected.

The limit as $\theta \rightarrow 0$ implies more counterintuitive phenomena. First,  $B$ and the rest of the system always remain in a tensor-product bipartite state—share no correlations, let alone entanglement. Yet  $B$ can still carry a nonclassically large amount of quantum Fisher information.  Furthermore, imagine, in addition to the $\theta \rightarrow 0$ limit, measuring $B$ between $t_1$ and $t_2$, before any other measurement and before $D$ is initialized.\footnote{In principle,   also the input state of the probe $B$ could be created at a later time and  probabilistically teleported to the beginning of the experiment.
} At time $t_5$, one would postprocess the data from the $B$ measurements. One would uncover the same contextuality as in conventional weak-value amplification [Fig. \ref{fig:CircuitRep}(a)]. This conclusion paradoxically holds even though $B$ is destroyed before $A$, $C$, and $D$ are measured. How? If PCTCs are real (perhaps probabilistic) effects of quantum theory, the nonclassicality comes from time travel.   Without real PCTCs, the paradox’s resolution will depend on  the power of entanglement.


Previous works have addressed the  advantages offered by CTCs  \cite{Brun03, Aaronson04, Aaronson09,  Brun09, Lloyd11, Brun12, Brun13, Pienaar13, Bub14,Bartkiewicz19}.  
For example,  PCTCs would boost a computer's computational power \cite{Brun03, Aaronson04, Aaronson09,  Lloyd11, Brun12}. (Classical computers, too, can achieve such computational power if postselected.) Our metrological protocol differs, posing a paradox even in the absence of true CTCs: Probabilistically simulating PCTCs suffices for achieving the nonclassical advantage. Relatedly, Svetlichny showed that PCTC simulation can effect a Bell measurement of a state before the state is created~\cite{Svetlichny11}. Also, probabilistically simulating DCTCs enables nonorthogonal-state discrimination  \cite{Vairogs22}. However, our result differs from these two by entailing that CTC simulation can effectively enable a truly nonclassical advantage---one sourced by contextuality---in the past.

\textit{Conclusions}.---We have shown how simulating time travel with entanglement benefits the estimation of a coupling strength.  A certain ``key'' input state is needed to unlock a quantum advantage.  However, in our setup, the ideal input state is  known only after the interaction takes place and the system is measured. We have shown how simulating quantum time travel allows for the key to be created at a later time and then effectively teleported back in time to serve as the experiment's input. The time travel can be simulated with postselected quantum-teleportation circuits.  Our Gedankenexperiment thus draws a metrological advantage from effective retrocausation founded in entangled states. 
While PCTC  simulations do not allow you to go back and alter your past, they do allow you to create a better tomorrow by fixing yesterday's problems today.

\medskip

\textit{Acknowledgements.---}The authors would like to thank Aharon Brodutch, Noah Lupu-Gladstein and Hugo Lepage for useful discussions. This work was supported by the  EPSRC, the Sweden-America Foundation, the Lars Hierta Memorial Foundation and Girton College. 

\newpage

\onecolumngrid

 \section*{ S\lowercase{upplementary} N\lowercase{ote} I:  F\lowercase{igures  2(a) and (b) prepare the same postselected $B$ state} }
 
 Here, we show that the postselected final state of $B$ in Fig. 2(a) equals that in Fig. 2(b).  In the main text, we showed that the postselected final state of $B$ in Fig. 2(a) is [Eq.  (4)]
\begin{align}
\ket{\Psi^{\rm{PS}} (\theta)}_{{B}} & = \frac{1}{\sqrt{\pps}}  \ket{\psi^{\rm{PS}} (\theta)}_{{B}}   \\
& \equiv \frac{1}{\sqrt{\pps}} \left( _{{A}}\bra{f} \otimes \hat{1}_{{B}} \right) e^{- i \theta \hat{\Pi}_{a} \otimes \hat{B} / 2} \ket{i}_{{A}} \ket{\psi}_B ,
\end{align}
where $p_{\theta}^{\mathrm{PS}}  \equiv {}_{B}\braket{\psi^{\rm{PS}}  (\theta)|\psi^{\rm{PS}}  (\theta)}_{{B}}$ and $\ket{\psi}_B \equiv \frac{1}{\sqrt{2} } ( \ket{b^+}_{{B}} + \ket{b^-}_{{B}} )$

We now calculate how the whole-system state evolves throughout our protocol, as described in the main text and Fig. 2(b).  Immediately after $t_0$, the joint state is
\begin{align}
\ket{\Psi (t_0)}_{A,B,C} & \equiv \frac{1}{\sqrt{2}} \left( \ket{0}_{{A}} \ket{0}_{{C}} +  \ket{1}_{{A}}\ket{1}_{{C}}\right) \ket{\psi}_B .
\end{align}
Immediately after  $t_1$, the joint state is
\begin{align}
\ket{\Psi_{\theta} (t_1)}_{A,B,C} & \equiv \left( e^{- i \theta \hat{\Pi}_{a} \otimes \hat{B} / 2} \otimes \hat{1}_C \right) \ket{\Psi (t_0)}_{A,B,C} .
\end{align}
Immediately after $t_2$,  $A$ is measured,  yielding an outcome $_A \bra{f^{\pm}}$. The joint state is
\begin{align}
\ket{\Psi_{\theta} (t_2)}_{B,C} & \equiv \frac{1}{ \sqrt{N_2}}  \left( _A \bra{f^{\pm}} \otimes \hat{1}_{B,C} \right) \left( e^{- i \theta \hat{\Pi}_{a} \otimes \hat{B} / 2} \otimes \hat{1}_C \right) \ket{\Psi (t_0)}_{A,B,C} .
\end{align}
$N_2$ is a normalization factor. 
 Immediately after  $t_3$, the joint state  is
\begin{align}
\ket{\Psi_{\theta} (t_3)}_{B,C,D} & \equiv  \ket{\Psi_{\theta} (t_2)}_{B,C} \ket{i^{\pm}}_D \\
& = \frac{1}{ \sqrt{N_2}}  \left( _A \bra{f^{\pm}} \otimes \hat{1}_{B,C,D} \right) \left( e^{- i \theta \hat{\Pi}_{a} \otimes \hat{B} / 2} \otimes \hat{1}_{C,D} \right)  \ket{\Psi (t_0)}_{A,B,C} \ket{i^{\pm}}_D .
\end{align}

At $t_4$ qubits  $C$ and  $D$ are measured in the Bell basis, $\left\{ \ket{\Psi^{\pm}} = \frac{1}{\sqrt{2}} \left( \ket{0}\ket{1}  \pm \ket{1}\ket{0} \right)  , \;   \ket{\Phi^{\pm}} = \frac{1}{\sqrt{2}} \left( \ket{0}\ket{0} \pm \ket{1}\ket{1} \right) \right\}$.  At $t_5$, $B$ is discarded unless the Bell measurement yielded outcome $_{{C},{D}} \bra{\Phi^+}$.  Immediately after $t_5$, the state is 
\begin{align}
\ket{\Psi_{\theta} (t_5)}_{B} & \equiv \frac{1}{ \sqrt{N_5}}  \left(\hat{1}_{B}  \otimes {_{{C},{D}} \bra{\Phi^+} } \right) \ket{\Psi (t_3)}_{B,C,D}  \\
& = \frac{1}{ \sqrt{N_5}}  \left( _A \bra{f^{\pm}} \otimes \hat{1}_{B} \otimes {_{{C},{D}} \bra{\Phi^+} } \right) \left( e^{- i \theta \hat{\Pi}_{a} \otimes \hat{B} / 2} \otimes \hat{1}_{C,D} \right)  \ket{\Psi (t_0)}_{A,B,C} \ket{i^{\pm}}_D  \\
& =  \frac{1}{  \sqrt{N_5}}  \left( _A \bra{f^{\pm}} \otimes \hat{1}_{B} \otimes {_{{C},{D}} \bra{\Phi^+} } \right) \left( e^{- i \theta \hat{\Pi}_{a} \otimes \hat{B} / 2} \otimes \hat{1}_{C,D} \right) \frac{1}{\sqrt{2}}
\left( \ket{0}_{{A}} \ket{0}_{{C}} +  \ket{1}_{{A}}\ket{1}_{{C}}\right) \ket{\psi}_B 
\ket{i^{\pm}}_D .
\end{align}
$N_5$ is a normalization factor. We can expand the factor $\ket{\Phi^+ }_{A,C} \ket{i^{\pm}}$ in  the Bell basis for $C$ and $D$:
\begin{align}
\ket{\Psi_{\theta} (t_5)}_{B} 
 = &  \frac{1}{  \sqrt{N_5}}  \left( _A \bra{f^{\pm}} \otimes \hat{1}_{B} \otimes {_{{C},{D}} \bra{\Phi^+} } \right) \left( e^{- i \theta \hat{\Pi}_{a} \otimes \hat{B} / 2} \otimes \hat{1}_{C,D} \right) \ket{\psi}_B   \nonumber \\
 & \times
\frac{1}{2} \left[ \ket{\Phi^+ }_{C,D} \ket{i^{\pm}}_A 
+  \ket{\Phi^- }_{C,D} \hat{Z}_A \ket{i^{\pm}}_A  + 
  \ket{\Psi^+ }_{C,D} \hat{X}_A \ket{i^{\pm}}_A   +  \ket{\Psi^- }_{C,D} \hat{X}_A \hat{Z}_A \ket{i^{\pm}}_A  
\right] .
\end{align}
$\hat{X}$ and $\hat{Z}$ denote  Pauli matrices.  We simplify the above expression by calculating the inner product with
 $_{{C},{D}} \bra{\Phi^+}$:
\begin{align}
\ket{\Psi_{\theta} (t_5)}_{B} 
  = &  \frac{1}{  \sqrt{N_5}}  \left( _A \bra{f^{\pm}} \otimes \hat{1}_{B}  \right) \left( e^{- i \theta \hat{\Pi}_{a} \otimes \hat{B} / 2} \right)    \ket{i^{\pm}}_A \ket{\psi}_B 
.
\end{align}
We choose the optimal system preparation $\ket{\psi}_B = \frac{1}{\sqrt{2} } ( \ket{b^+}_{{B}} + \ket{b^-}_{{B}} )$.   Explicitly calculating the normalization factor yields $N_5 = \pps$.  Consequently,  $\ket{\Psi_{\theta} (t_5)}_{B} = \ket{\Psi^{\rm{PS}} (\theta)}_{{B}}$, and the circuits preparing the postselected states in Figs. 2(a) and 2(b) are equivalent.

 \section*{ S\lowercase{upplementary} N\lowercase{ote} II:    S\lowercase{ketch of optical implementation}}

 Here, we provide a proposal for implementing our Gedankenexperiment using path-entangled photons and linear-optical operations.  The experimental layout is shown in Fig. \ref{fig:Optics}.

In our proposed optics realization,  the four qubits of Fig. 2(a) are encoded in two photons' polarization and path degrees of freedom.  $A$ and $B$ are respectively encoded in the path and polarization degrees of freedom of photon $2$. Similarly,  $C$ and $D$ are respectively encoded in the path and polarization degrees of freedom of photon $1$.  To summarize: $\ket{0/1}_A \equiv \ket{\downarrow / \uparrow}^{(2)}$,  $\ket{0/1}_C \equiv \ket{\downarrow / \uparrow}^{(1)}$,  $\ket{0/1}_B \equiv \ket{H / V}^{(2)}$, and  $\ket{0/1}_D \equiv \ket{H / V}^{(1)}$.   Here, $\{ \downarrow, \; \uparrow \}$ and $\{H, \; V \}$ denote the path and polarization bases,  respectively.

\renewcommand{\thefigure}{S1}
\begin{figure}
\includegraphics[scale=0.25]{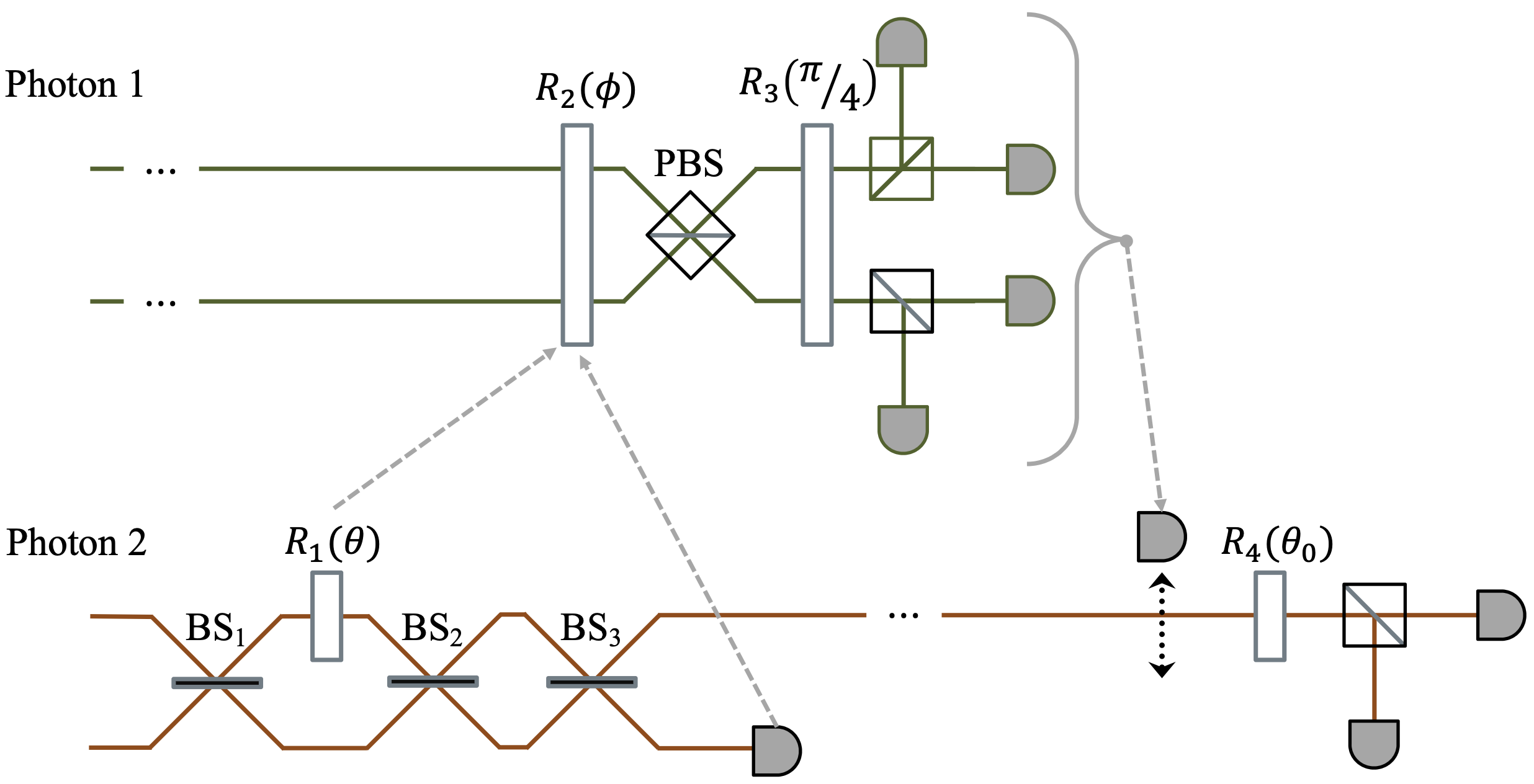}
\caption{Optical realization of the Gedankenexperiment shown in Fig. 2(b) of the main text.  $\mathrm{BS}_i$ denotes the $i\mathrm{th}$ beam-splitter, which implements the unitary evolution $\mathcal{B}_i$. $\mathrm{PBS}$ denotes polarizing beam-splitters. Detectors are shown in gray.  The experiment is described in more detail in the text. }
\label{fig:Optics}
\end{figure}

The experiment begins with two path-entangled photons of horizontal polarization.   We prepare this state as follows. References \cite{Barbieri05, Cinelli05, Barreiro05, Vallone07, Chen07, Ciampini16} show how spontaneous parametric down-conversion can be used to create the  two-photon state  $\frac{1}{2} \big( \ket{H}^{(1)}\ket{H}^{(2)}+\ket{V}^{(1)}\ket{V}^{(2)} \big) \big( \ket{\downarrow}^{(1)}\ket{\uparrow}^{(2)}+\ket{\uparrow}^{(1)}\ket{\downarrow}^{(2)} \big) $. Postselecting photons 1 and 2   on horizontal polarization, we obtain $\frac{1}{\sqrt{2}}  \ket{H}^{(1)}\ket{H}^{(2)} \big( \ket{\downarrow}^{(1)}\ket{\uparrow}^{(2)}+\ket{\uparrow}^{(1)}\ket{\downarrow}^{(2)} \big) $.  This state serves as the input  to the setup shown in Fig. \ref{fig:Optics}.

In the protocol's second step, we implement the unknown unitary $U=e^{- i \theta \hat{\Pi}_a \otimes \hat{B}/2}$, which rotates the polarization of photon $2$ conditionally on the photon's path state.  We set $\hat{B} = \hat{\sigma}_x$, such that the optimal state of  $B$ is $\ket{0}_B = \ket{H}^{(2)}$. Accordingly, $R_1(\theta)=e^{-i \theta \hat{\Pi}^{(2)}_{\uparrow} \otimes \hat{\sigma}_x^{(2)}/2}$ rotates the polarization of photon $2$ through an angle $\theta$ if the photon is in the upper path: $\ket{\uparrow}^{(2)}$. Beam-splitters $1$ and $2$ are tuned to rotate the path of photon $2$ such that $\mathcal{B}_2 \cdot  \hat{\Pi}^{(2)}_{\uparrow}  \cdot \mathcal{B}_1 = \hat{\Pi}_a$.

In the third step,   beam-splitter $3$ rotates the path degree of freedom of photon $2$ such that the following detector measures the photon in the $\{ \ket{f^{\pm} }_A \}$ basis.  (In practice,  beam-splitters $2$ and $3$ can be combined into one beam-splitter.)

In the fourth step,  two pieces of information are revealed: the $A$-measurement outcome and the form of $\hat{\Pi}_a$.  The revelation allows us to set the polarization rotator's angle $\phi$  such that $R_2(\phi)$ prepares photon 1's polarization (qubit $D$) in the optimal input state, $\ket{i^{\pm}}_D$.

In the fifth step,  $C$ and $D$ (photon $1$) are subject to a polarization-path Bell measurement, implemented in three steps. First,  the polarizing beam-splitter implements a CNOT gate with the polarization qubit as the control. Second,  the polarization rotator $R_2(\pi/4)$ implements a Hadamard gate on the polarization.  Third,  measurements of the $\{\ket{0/1}_{C} , \; \ket{0/1}_{D}  \}$ basis yield the required Bell measurement. Conditionally on measuring photon $1$ in the horizontal polarization and in its lower path,  the polarization state  ($\ket{i^{\pm}}_D$) of particle $1$  effectively travels backward in time and becomes the path state (qubit $A$)  of particle $2$.  Otherwise, the blocker destroys photon $2$.
Together, these four steps implement the quantum circuit shown in Fig.  2(b).

 \section*{S\lowercase{upplementary} N\lowercase{ote} III: B\lowercase{oosting the probability of preparing the optimal probe state}}

Our Gedankenexperiment's success probability is half the success probability of standard weak-value amplification.  However, one can double our success probability, at the cost of slightly weakening the connection with PCTCs.  One achieves this result by removing the blocker if the measurement at $t_5$ yields either $_{{C},{D}} \bra{\Phi^{+}}$ or $_{{C},{D}} \bra{\Psi^{+}}$.  First, we calculate  the  probability with which the circuits in Figs. 2(a) and 2(b) successfully prepare   $ \ket{\Psi^{\rm{PS}} (\theta)}_{{B}}$.  Then, we show how to boost our Gedankenexperiment's success probability.

Let us show that the main text's protocol succeeds with half the probability of standard weak-value amplification. The circuit in Fig. 2(a)  succeeds if the $A$ measurement yields the $\bra{f}$ outcome. As $A$ was prepared in $\ket{i}_A \equiv \ket{i^+}_A$, the success probability is $\pps = |\braket{f|i}|^2$.  The circuit in Fig. 2(b) succeeds if measuring $C$ and $D$ yields the $ _{C,D} \bra{\Phi^+}$ outcome. The state of  $CD$ depends on whether  $A$ ended up in  $_A \bra{f^+} \equiv _A \bra{f}$ or in  $_A \bra{f^-}$.  The total success probability is thus
\begin{align}
p_{\theta}^{\mathrm{PCTC}}  & = p( _A \bra{f^+}) \times  | _D \bra{i^+} _C \bra{f^+} \cdot \ket{\Phi^+}_{C,D}|^2 + p( _A \bra{f^-}) \times  | _D \bra{i^-} _C \bra{f^-} \cdot \ket{\Phi^+}_{C,D}|^2 \\
& = p( _A \bra{f^+}) \times  \frac{1}{2} | _D \braket{i^+ | f^+} _D |^2 + p( _A \bra{f^-}) \times \frac{1}{2}  | _D \braket{i^- | f^-} _D |^2 \\
& = \left[ p( _A \bra{f^+}) + p( _A \bra{f^-}) \right] \times  \frac{1}{2} | _D \braket{i^+ | f^+} _D |^2 
\\
& =   \frac{1}{2} | _D \braket{i^+ | f^+} _D |^2 
\\
& =   \frac{1}{2} \pps .
\end{align}
We applied $\ket{\Phi^+} = \frac{1}{\sqrt{2}} \left( \ket{f^{\pm}}\ket{f^{\pm}} + \ket{f^{\mp}}\ket{f^{\mp}} \right)$ in the second equality,   $|  \braket{i^+ | f^+}  |^2 = |  \braket{i^- | f^-}  |^2$ in the third equality, and  $p( _A \bra{f^+}) + p( _A  \bra{f^-}) = 1$ in the fourth equality. Consequently, the success probability of the circuit in Fig. 2(b) is half the success probability of standard weak-value amplification [Fig. 2(a)]: $p_{\theta}^{\mathrm{PCTC}}/ \pps = \frac{1}{2}$.

Our Gedankenexperiment's  success probability  can be boosted by a factor of $2$.  To achieve this boost, one postselects  also on the $CD$ state's being measured in  $ _{C,D} \bra{\Psi^+}$. (In the original Gedankenexperiment, one postselects  only on the $ _{C,D}  \bra{\Phi^+}$ outcome.)  We prove these claims now.  One can, without loss of generality,  orient the $z$-axis such that the $\hat{\sigma}_z$ eigenbasis is the $\hat{\Pi}_a$ eigenbasis.  Thus, to simplify the proof, we set these eigenbases equal.  Further, we shift the eigenvalues ($1$ and $0$) of $\hat{\Pi}_a$ by $-\frac{1}{2}$ such that $\hat{\Pi}_a = \hat{\sigma}_z/2$.  This shift   simplifies the algebra that now follows.\footnote{It is possible to boost also  the success probability of an experiment where $\hat{\Pi}_a = \ket{1}\bra{1}$ (i.e., no shift). However, the algebra describing that procedure is tedious and less illuminating. }

If $\hat{\Pi}_a = \hat{\sigma}_z/2$, then the optimal $A$-measurement basis  is $ \{ \ket{f^{\pm}} \} = \{ \ket{\pm} \}$.  We set the input state of  $D$ to $\ket{i^{\pm}}_D = \cos(\gamma) \ket{0}_D \pm \sin(\gamma) \ket{1}_D $. This input ensures that the two possible weak values are equal:
\begin{equation}
 _{f^+}\braket{\hat{\Pi}_{a}}_{i^+}  =  _{f^-}\braket{\hat{\Pi}_{a}}_{i^-}  = \frac{1}{2} \frac{\cos(\gamma) - \sin(\gamma)}{\cos(\gamma) + \sin(\gamma)}  .
\end{equation}
$ _{f^{\pm}}\braket{\hat{\Pi}_{a}}_{i^{\pm}}$ is anomalous (i.e, $>1$) if $\frac{3 \pi}{4} - \mathrm{arccot}\left( 2 \right) < \gamma < \frac{3 \pi}{4} + \mathrm{arccot}\left( 2 \right)$.
As we have seen above, if the $ _{C,D} \bra{\Phi^+}$  outcome is recorded, the state $\ket{i^{\pm}}_D$ is effectively teleported to the input of  $A$ such that the metrologically advantageous state $  e^{- i \theta  _{f^+}\braket{\hat{\Pi}_{a}}_{i^+}  \hat{B} / 2}     \ket{\psi}_B $ is prepared.  A routine calculation shows that  the $ _{C,D} \bra{\Psi^+}$ outcome occurs with the same probability ($ \frac{1}{2} | _D \braket{i^+ | f^+} _D |^2$) as the $ _{C,D} \bra{\Phi^+}$ outcome. But if the $ _{C,D} \bra{\Psi^+} $ outcome is recorded,  $A$ is effectively prepared in the state  $\ket{j^{\pm}}_A = \sin(\gamma) \ket{0}_A \pm \cos(\gamma) \ket{1}_A$.  These two states result in two equal weak values:
\begin{equation}
 _{f^+}\braket{\hat{\Pi}_{a}}_{j^+}  =  _{f^-}\braket{\hat{\Pi}_{a}}_{j^-}  = - \frac{1}{2}  \frac{\cos(\gamma) - \sin(\gamma)}{\cos(\gamma)+ \sin(\gamma)} = - \left( _{f^{\pm}}\braket{\hat{\Pi}_{a}}_{i^{\pm}} \right) .
\end{equation}
The $ _{C,D} \bra{\Psi^+} $ outcome  prepares the  state  $ e^{ i \theta  _{f^+}\braket{\hat{\Pi}_{a}}_{i^+}  \hat{B} / 2} \ket{\psi}_B $. Only the sign in the exponent distinguishes this state from the state prepared when the  $ _{C,D} \bra{\Phi^+} $ outcome is recorded. This sign is insignificant for metrological purposes.  
By Eq. (7), the two states have the same quantum Fisher information.

To summarize,  the probability of successfully preparing the optimal input state, using the circuit in Fig. 2(b),  is half the success probability of standard weak-value amplification [Fig. 2(a)].  However, by postselecting the Fig. 2(a) circuit on \textit{both} the $_{C,D} \bra{\Phi^+} $ and $_{C,D} \bra{\Psi^+}$ outcomes,  we render the success probabilities equal.

 \section*{S\lowercase{upplementary} N\lowercase{ote} IV:   P\lowercase{reparation of versatile metrology probes via effective retrocausality}}

We propose a  protocol for leveraging effective quantum retrocausality in a more practical metrological scenario.   Consider estimating the strength $\theta$ of the single-qubit unitary $U_{\vec{n}}(\theta)=e^{-i \theta \vec{\sigma} \cdot \vec{n}/2}$.  $\vec{\sigma}$ denotes the Pauli vector,  and $ \vec{n}$ denotes a three-dimensional unit vector.  The optimal probe states are  $ \ket{\vec{n}^{\pm}} \equiv \frac{1}{\sqrt{2}} \left(\ket{\vec{n}^0} \pm \ket{\vec{n}^1} \right)$,  which maximize the quantum Fisher information: $\max_{\ket{\psi_0}} \left\{ \mathcal{I}_{\mathrm{q}}\bm{(}\theta | U_{\vec{n}}(\theta) \ket{\psi_0} \bm{)} \right\} =\mathcal{I}_{\mathrm{q}}\bm{(}\theta | U_{\vec{n}}(\theta) \ket{\vec{n}^{\pm}} \bm{)} = 1$.   $\ket{\vec{n}^0}$ and $\ket{\vec{n}^1}$ denote the eigenstates  of $\vec{\sigma} \cdot \vec{n}$ and correspond to the eigenvalues $-1$ and $+1$, respectively.  
 It might appear impossible to prepare a probe that will be optimal  for all $\vec{n}$.  Similarly,  it might appear impossible to prepare an optimal probe if $\vec{n}$ is known only after the interaction.  However, entanglement manipulation can be used to effectively initialize optimal probes after the unitary is performed.

In our protocol, one prepares a singlet Bell state  $\ket{\Psi^{-}}_{{A},{B}}= \frac{1}{\sqrt{2}} \left( \ket{\vec{n}^+}_{{A}} \ket{\vec{n}^-}_{{B}} -  \ket{\vec{n}^-}_{{A}}\ket{\vec{n}^+}_{{B}}\right)$, whose form is independent of $\vec{n}$. The  $A$ qubit is subject to $U_{\vec{n}}(\theta)$.  After the interaction, when  $\vec{n}$ is known,   one measures the $B$ qubit's $\{  \ket{\vec{n}^{\pm}} \}$ basis.  The measurement effectively teleports to the past an optimal input state for the probe qubit $A$.    
This protocol is not directly related to PCTCs. However, it is a more practical and diverse metrological protocol than the Gedankenexperiment discussed in the main text: the present protocol    effectively teleports useful states to the beginning of standard metrology experiments.  Below, we detail this protocol's workings, depicted in Fig. \ref{fig:RetroMet}(b).

 \renewcommand{\thefigure}{S2}
\begin{figure}
\includegraphics[scale=0.25]{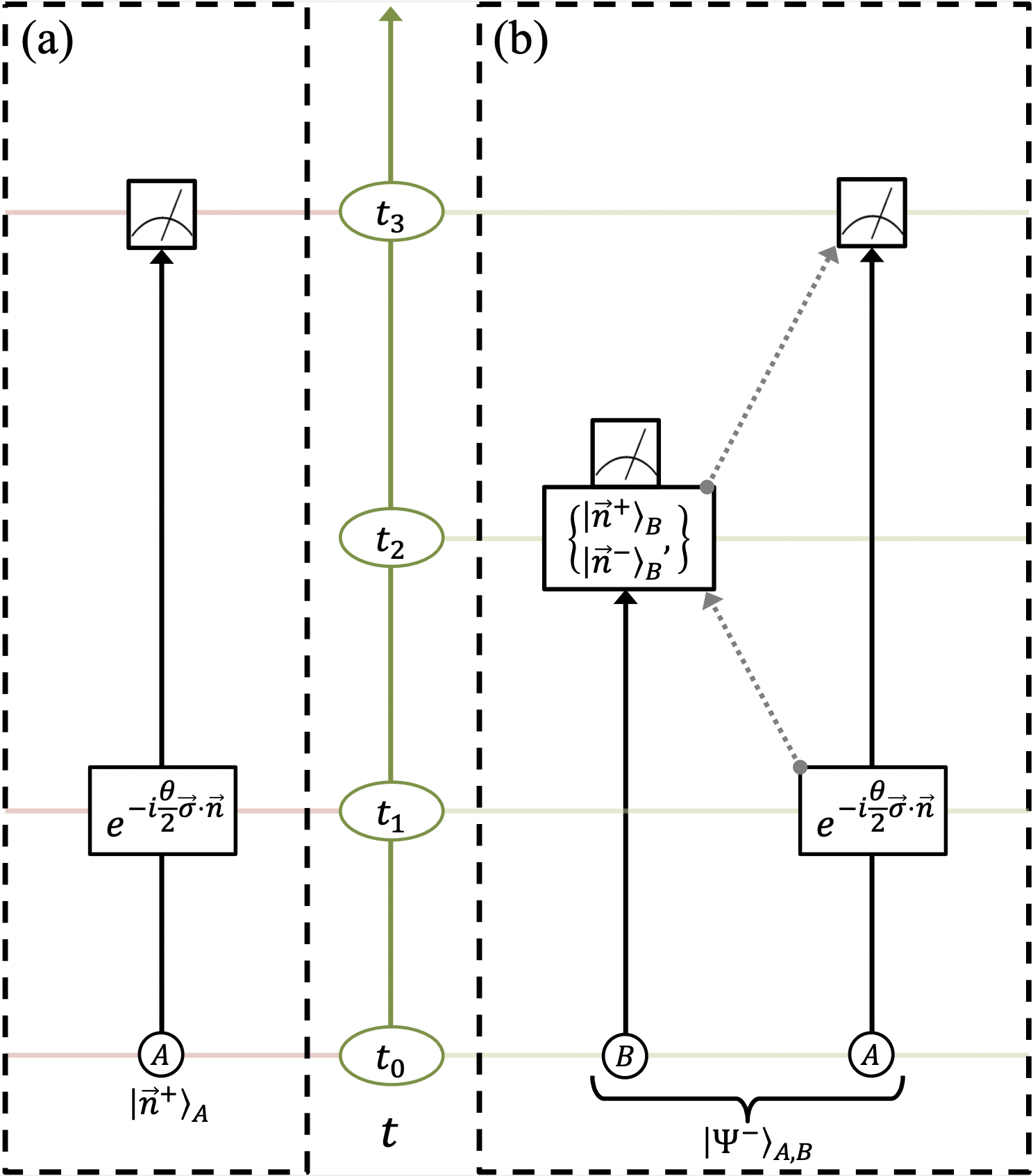}
\caption{Circuit diagrams for single-qubit metrology with (a) a standard setup, and (b) a setup that leverages effective retrocausality. Time progresses in the laboratory's rest frame as one proceeds upward along the central, vertical axis. Black lines represent qubits. Dashed gray lines represent classical information.}
\label{fig:RetroMet}
\end{figure}

Figure \ref{fig:RetroMet} shows two quantum circuits for measuring the strength $\theta$ of an unknown interaction $U_{\vec{n}}(\theta)=e^{-i \theta \vec{\sigma} \cdot \vec{n}/2}$.  The circuit in Fig. \ref{fig:RetroMet}(a) relies on \textit{a priori} knowledge of $\vec{n}$.
If one wishes to prepare a probe that is optimal for any $U_{\vec{n}}(\theta)$, in an $\vec{n}$-agnostic fashion, one can use the circuit shown in Fig. \ref{fig:RetroMet}(b).  The circuit involves two qubits, $A$ and $B$.  The protocol proceeds as follows:
\begin{enumerate}
\item[$t_0$: ]{
$\bullet$
$A$ and $B$ are entangled: $\ket{\Psi^{-}}_{{A},{B}}= \frac{1}{\sqrt{2}} \left( \ket{0}_{{A}} \ket{1}_{{B}} -  \ket{1}_{{A}}\ket{0}_{{B}}\right) =\frac{1}{\sqrt{2}} \left( \ket{\vec{n}^+}_{{A}} \ket{\vec{n}^-}_{{B}} - \ket{\vec{n}^-}_{{A}}\ket{\vec{n}^+}_{{B}}\right) $.
}
\item[$t_1$: ]{$\bullet$ $A$ is subject to the unitary $U_{\vec{n}}(\theta)=e^{-i \theta \vec{\sigma} \cdot \vec{n}/2}$. \\
$\bullet$ The joint state becomes  $ \left( e^{-i \theta \vec{\sigma} \cdot \vec{n} /2} \otimes \hat{1}_B \right) \ket{\Psi^{-}}_{{A},{B}} $.
}
\item[$t_2$: ]{
$\bullet$ The  form of $\vec{n}$ is now known.  \\
$\bullet$ $B$ is measured in the $\{ \ket{\vec{n}^{\pm}} \}$ basis. \\
$\bullet$ Depending on the outcome, the state of $A$ becomes $$ \sqrt{2} \left( \hat{1} \otimes _B \bra{{\vec{n}^{\pm}}} \right) \left(e^{-i \theta \vec{\sigma} \cdot \vec{n}} \otimes \hat{1}_B \right) \ket{\Psi^{-}}_{{A},{B}} =  \sqrt{2} e^{-i \theta \vec{\sigma} \cdot \vec{n} /2} \left( \hat{1} \otimes _B \bra{{\vec{n}^{\pm}}} \right)  \ket{\Psi^{-}}_{{A},{B}} = e^{-i \theta \vec{\sigma} \cdot \vec{n}/2} \ket{\vec{n}^{\mp}} .$$ This is a metrologically optimal state.
}
\item[$t_3$: ]{$\bullet$ $A$ is measured in the optimal $\{ \ket{\vec{n}^{\pm}} \}$ basis. }
\end{enumerate}

To estimate $\theta$, one would use our protocol to prepare and measure several metrologically optimal $A$ qubits. One could do so in parallel.
Thus,  the circuit in Fig. \ref{fig:RetroMet}(b) effectively enables the  preparation of an optimal input state to the unknown unitary, after the unitary has occurred. In contrast with weak-value amplification and our PCTC Gedankenexperiment, the success probability is one.   Moreover, the protocol depicted in  Fig. \ref{fig:RetroMet}(b) leverages retrocausality in a way that can be useful for general metrology schemes—not only postselected schemes. Although we focused on a qubit implementation,   generalizations to higher dimensions are straightforward.

\bibliography{CTCWeak}

\begin{thebibliography}{61}%
\makeatletter
\providecommand \@ifxundefined [1]{%
 \@ifx{#1\undefined}
}%
\providecommand \@ifnum [1]{%
 \ifnum #1\expandafter \@firstoftwo
 \else \expandafter \@secondoftwo
 \fi
}%
\providecommand \@ifx [1]{%
 \ifx #1\expandafter \@firstoftwo
 \else \expandafter \@secondoftwo
 \fi
}%
\providecommand \natexlab [1]{#1}%
\providecommand \enquote  [1]{``#1''}%
\providecommand \bibnamefont  [1]{#1}%
\providecommand \bibfnamefont [1]{#1}%
\providecommand \citenamefont [1]{#1}%
\providecommand \href@noop [0]{\@secondoftwo}%
\providecommand \href [0]{\begingroup \@sanitize@url \@href}%
\providecommand \@href[1]{\@@startlink{#1}\@@href}%
\providecommand \@@href[1]{\endgroup#1\@@endlink}%
\providecommand \@sanitize@url [0]{\catcode `\\12\catcode `\$12\catcode
  `\&12\catcode `\#12\catcode `\^12\catcode `\_12\catcode `\%12\relax}%
\providecommand \@@startlink[1]{}%
\providecommand \@@endlink[0]{}%
\providecommand \url  [0]{\begingroup\@sanitize@url \@url }%
\providecommand \@url [1]{\endgroup\@href {#1}{\urlprefix }}%
\providecommand \urlprefix  [0]{URL }%
\providecommand \Eprint [0]{\href }%
\providecommand \doibase [0]{http://dx.doi.org/}%
\providecommand \selectlanguage [0]{\@gobble}%
\providecommand \bibinfo  [0]{\@secondoftwo}%
\providecommand \bibfield  [0]{\@secondoftwo}%
\providecommand \translation [1]{[#1]}%
\providecommand \BibitemOpen [0]{}%
\providecommand \bibitemStop [0]{}%
\providecommand \bibitemNoStop [0]{.\EOS\space}%
\providecommand \EOS [0]{\spacefactor3000\relax}%
\providecommand \BibitemShut  [1]{\csname bibitem#1\endcsname}%
\let\auto@bib@innerbib\@empty
\bibitem [{\citenamefont {Giovannetti}\ \emph {et~al.}(2011)\citenamefont
  {Giovannetti}, \citenamefont {Lloyd},\ and\ \citenamefont
  {Maccone}}]{Giovanetti11}%
  \BibitemOpen
  \bibfield  {author} {\bibinfo {author} {\bibfnamefont {V.}~\bibnamefont
  {Giovannetti}}, \bibinfo {author} {\bibfnamefont {S.}~\bibnamefont {Lloyd}},
  \ and\ \bibinfo {author} {\bibfnamefont {L.}~\bibnamefont {Maccone}},\
  }\href@noop {} {\bibfield  {journal} {\bibinfo  {journal} {Nature photonics}\
  }\textbf {\bibinfo {volume} {5}},\ \bibinfo {pages} {222} (\bibinfo {year}
  {2011})}\BibitemShut {NoStop}%
\bibitem [{\citenamefont {Salmon}\ \emph {et~al.}(2022)\citenamefont {Salmon},
  \citenamefont {Strelchuk},\ and\ \citenamefont
  {Arvidsson-Shukur}}]{Salmon22}%
  \BibitemOpen
  \bibfield  {author} {\bibinfo {author} {\bibfnamefont {W.}~\bibnamefont
  {Salmon}}, \bibinfo {author} {\bibfnamefont {S.}~\bibnamefont {Strelchuk}}, \
  and\ \bibinfo {author} {\bibfnamefont {D.}~\bibnamefont {Arvidsson-Shukur}},\
  }\href {\doibase 10.48550/ARXIV.2205.14142} {\  (\bibinfo {year} {2022}),\
  10.48550/ARXIV.2205.14142}\BibitemShut {NoStop}%
\bibitem [{\citenamefont {Jordan}\ \emph {et~al.}(2014)\citenamefont {Jordan},
  \citenamefont {Mart{\'\i}nez-Rinc{\'o}n},\ and\ \citenamefont
  {Howell}}]{Jordan14}%
  \BibitemOpen
  \bibfield  {author} {\bibinfo {author} {\bibfnamefont {A.~N.}\ \bibnamefont
  {Jordan}}, \bibinfo {author} {\bibfnamefont {J.}~\bibnamefont
  {Mart{\'\i}nez-Rinc{\'o}n}}, \ and\ \bibinfo {author} {\bibfnamefont {J.~C.}\
  \bibnamefont {Howell}},\ }\href@noop {} {\bibfield  {journal} {\bibinfo
  {journal} {Physical Review X}\ }\textbf {\bibinfo {volume} {4}},\ \bibinfo
  {pages} {011031} (\bibinfo {year} {2014})}\BibitemShut {NoStop}%
\bibitem [{\citenamefont {Dressel}\ \emph {et~al.}(2014)\citenamefont
  {Dressel}, \citenamefont {Malik}, \citenamefont {Miatto}, \citenamefont
  {Jordan},\ and\ \citenamefont {Boyd}}]{Dressel14}%
  \BibitemOpen
  \bibfield  {author} {\bibinfo {author} {\bibfnamefont {J.}~\bibnamefont
  {Dressel}}, \bibinfo {author} {\bibfnamefont {M.}~\bibnamefont {Malik}},
  \bibinfo {author} {\bibfnamefont {F.~M.}\ \bibnamefont {Miatto}}, \bibinfo
  {author} {\bibfnamefont {A.~N.}\ \bibnamefont {Jordan}}, \ and\ \bibinfo
  {author} {\bibfnamefont {R.~W.}\ \bibnamefont {Boyd}},\ }\href {\doibase
  10.1103/RevModPhys.86.307} {\bibfield  {journal} {\bibinfo  {journal} {Rev.
  Mod. Phys.}\ }\textbf {\bibinfo {volume} {86}},\ \bibinfo {pages} {307}
  (\bibinfo {year} {2014})}\BibitemShut {NoStop}%
\bibitem [{\citenamefont {Harris}\ \emph {et~al.}(2017)\citenamefont {Harris},
  \citenamefont {Boyd},\ and\ \citenamefont {Lundeen}}]{Harris17}%
  \BibitemOpen
  \bibfield  {author} {\bibinfo {author} {\bibfnamefont {J.}~\bibnamefont
  {Harris}}, \bibinfo {author} {\bibfnamefont {R.~W.}\ \bibnamefont {Boyd}}, \
  and\ \bibinfo {author} {\bibfnamefont {J.~S.}\ \bibnamefont {Lundeen}},\
  }\href@noop {} {\bibfield  {journal} {\bibinfo  {journal} {Phys. Rev. Lett.}\
  }\textbf {\bibinfo {volume} {118}},\ \bibinfo {pages} {070802} (\bibinfo
  {year} {2017})}\BibitemShut {NoStop}%
\bibitem [{\citenamefont {Arvidsson-Shukur}\ \emph {et~al.}(2020)\citenamefont
  {Arvidsson-Shukur}, \citenamefont {Yunger~Halpern}, \citenamefont {Lepage},
  \citenamefont {Lasek}, \citenamefont {Barnes},\ and\ \citenamefont
  {Lloyd}}]{ArvShukur19-2}%
  \BibitemOpen
  \bibfield  {author} {\bibinfo {author} {\bibfnamefont {D.~R.~M.}\
  \bibnamefont {Arvidsson-Shukur}}, \bibinfo {author} {\bibfnamefont
  {N.}~\bibnamefont {Yunger~Halpern}}, \bibinfo {author} {\bibfnamefont
  {H.~V.}\ \bibnamefont {Lepage}}, \bibinfo {author} {\bibfnamefont {A.~A.}\
  \bibnamefont {Lasek}}, \bibinfo {author} {\bibfnamefont {C.~H.~W.}\
  \bibnamefont {Barnes}}, \ and\ \bibinfo {author} {\bibfnamefont
  {S.}~\bibnamefont {Lloyd}},\ }\href {\doibase 10.1038/s41467-020-17559-w}
  {\bibfield  {journal} {\bibinfo  {journal} {Nature Communications}\ }\textbf
  {\bibinfo {volume} {11}},\ \bibinfo {pages} {3775} (\bibinfo {year}
  {2020})}\BibitemShut {NoStop}%
\bibitem [{\citenamefont {Lupu-Gladstein}\ \emph {et~al.}(2022)\citenamefont
  {Lupu-Gladstein}, \citenamefont {Yilmaz}, \citenamefont {Arvidsson-Shukur},
  \citenamefont {Brodutch}, \citenamefont {Pang}, \citenamefont {Steinberg},\
  and\ \citenamefont {Halpern}}]{Lupu22}%
  \BibitemOpen
  \bibfield  {author} {\bibinfo {author} {\bibfnamefont {N.}~\bibnamefont
  {Lupu-Gladstein}}, \bibinfo {author} {\bibfnamefont {Y.~B.}\ \bibnamefont
  {Yilmaz}}, \bibinfo {author} {\bibfnamefont {D.~R.~M.}\ \bibnamefont
  {Arvidsson-Shukur}}, \bibinfo {author} {\bibfnamefont {A.}~\bibnamefont
  {Brodutch}}, \bibinfo {author} {\bibfnamefont {A.~O.~T.}\ \bibnamefont
  {Pang}}, \bibinfo {author} {\bibfnamefont {A.~M.}\ \bibnamefont {Steinberg}},
  \ and\ \bibinfo {author} {\bibfnamefont {N.~Y.}\ \bibnamefont {Halpern}},\
  }\href {\doibase 10.1103/PhysRevLett.128.220504} {\bibfield  {journal}
  {\bibinfo  {journal} {Phys. Rev. Lett.}\ }\textbf {\bibinfo {volume} {128}},\
  \bibinfo {pages} {220504} (\bibinfo {year} {2022})}\BibitemShut {NoStop}%
\bibitem [{\citenamefont {Semenov}\ \emph {et~al.}(2023)\citenamefont
  {Semenov}, \citenamefont {Samelin}, \citenamefont {Boldt}, \citenamefont
  {Schünemann}, \citenamefont {Reiher}, \citenamefont {Vogel},\ and\
  \citenamefont {Hage}}]{Semenov23}%
  \BibitemOpen
  \bibfield  {author} {\bibinfo {author} {\bibfnamefont {A.~A.}\ \bibnamefont
  {Semenov}}, \bibinfo {author} {\bibfnamefont {J.}~\bibnamefont {Samelin}},
  \bibinfo {author} {\bibfnamefont {C.}~\bibnamefont {Boldt}}, \bibinfo
  {author} {\bibfnamefont {M.}~\bibnamefont {Schünemann}}, \bibinfo {author}
  {\bibfnamefont {C.}~\bibnamefont {Reiher}}, \bibinfo {author} {\bibfnamefont
  {W.}~\bibnamefont {Vogel}}, \ and\ \bibinfo {author} {\bibfnamefont
  {B.}~\bibnamefont {Hage}},\ }\href@noop {} {\  (\bibinfo {year} {2023})},\
  \Eprint {http://arxiv.org/abs/2303.14246} {arXiv:2303.14246 [quant-ph]}
  \BibitemShut {NoStop}%
\bibitem [{\citenamefont {Jenne}\ and\ \citenamefont
  {Arvidsson-Shukur}(2021)}]{Jenne21}%
  \BibitemOpen
  \bibfield  {author} {\bibinfo {author} {\bibfnamefont {J.~H.}\ \bibnamefont
  {Jenne}}\ and\ \bibinfo {author} {\bibfnamefont {D.~R.~M.}\ \bibnamefont
  {Arvidsson-Shukur}},\ }\href {\doibase 10.48550/ARXIV.2104.09520} {\
  (\bibinfo {year} {2021}),\ 10.48550/ARXIV.2104.09520}\BibitemShut {NoStop}%
\bibitem [{\citenamefont {Scandi}\ \emph {et~al.}(2023)\citenamefont {Scandi},
  \citenamefont {Abiuso}, \citenamefont {Surace},\ and\ \citenamefont
  {Santis}}]{Scandi23}%
  \BibitemOpen
  \bibfield  {author} {\bibinfo {author} {\bibfnamefont {M.}~\bibnamefont
  {Scandi}}, \bibinfo {author} {\bibfnamefont {P.}~\bibnamefont {Abiuso}},
  \bibinfo {author} {\bibfnamefont {J.}~\bibnamefont {Surace}}, \ and\ \bibinfo
  {author} {\bibfnamefont {D.~D.}\ \bibnamefont {Santis}},\ }\href@noop {} {\
  (\bibinfo {year} {2023})},\ \Eprint {http://arxiv.org/abs/2304.14984}
  {arXiv:2304.14984 [quant-ph]} \BibitemShut {NoStop}%
\bibitem [{\citenamefont {Pang}\ \emph {et~al.}(2014)\citenamefont {Pang},
  \citenamefont {Dressel},\ and\ \citenamefont {Brun}}]{Pang14}%
  \BibitemOpen
  \bibfield  {author} {\bibinfo {author} {\bibfnamefont {S.}~\bibnamefont
  {Pang}}, \bibinfo {author} {\bibfnamefont {J.}~\bibnamefont {Dressel}}, \
  and\ \bibinfo {author} {\bibfnamefont {T.~A.}\ \bibnamefont {Brun}},\ }\href
  {\doibase 10.1103/PhysRevLett.113.030401} {\bibfield  {journal} {\bibinfo
  {journal} {Phys. Rev. Lett.}\ }\textbf {\bibinfo {volume} {113}},\ \bibinfo
  {pages} {030401} (\bibinfo {year} {2014})}\BibitemShut {NoStop}%
\bibitem [{\citenamefont {Xu}\ \emph {et~al.}(2020)\citenamefont {Xu},
  \citenamefont {Liu}, \citenamefont {Datta}, \citenamefont {Knee},
  \citenamefont {Lundeen}, \citenamefont {Lu},\ and\ \citenamefont
  {Zhang}}]{Xu20}%
  \BibitemOpen
  \bibfield  {author} {\bibinfo {author} {\bibfnamefont {L.}~\bibnamefont
  {Xu}}, \bibinfo {author} {\bibfnamefont {Z.}~\bibnamefont {Liu}}, \bibinfo
  {author} {\bibfnamefont {A.}~\bibnamefont {Datta}}, \bibinfo {author}
  {\bibfnamefont {G.~C.}\ \bibnamefont {Knee}}, \bibinfo {author}
  {\bibfnamefont {J.~S.}\ \bibnamefont {Lundeen}}, \bibinfo {author}
  {\bibfnamefont {Y.-q.}\ \bibnamefont {Lu}}, \ and\ \bibinfo {author}
  {\bibfnamefont {L.}~\bibnamefont {Zhang}},\ }\href {\doibase
  10.1103/PhysRevLett.125.080501} {\bibfield  {journal} {\bibinfo  {journal}
  {Phys. Rev. Lett.}\ }\textbf {\bibinfo {volume} {125}},\ \bibinfo {pages}
  {080501} (\bibinfo {year} {2020})}\BibitemShut {NoStop}%
\bibitem [{\citenamefont {Jenne}\ and\ \citenamefont
  {Arvidsson-Shukur}(2022)}]{ArvShukur21}%
  \BibitemOpen
  \bibfield  {author} {\bibinfo {author} {\bibfnamefont {J.~H.}\ \bibnamefont
  {Jenne}}\ and\ \bibinfo {author} {\bibfnamefont {D.~R.~M.}\ \bibnamefont
  {Arvidsson-Shukur}},\ }\href {\doibase 10.1103/PhysRevA.106.042404}
  {\bibfield  {journal} {\bibinfo  {journal} {Phys. Rev. A}\ }\textbf {\bibinfo
  {volume} {106}},\ \bibinfo {pages} {042404} (\bibinfo {year}
  {2022})}\BibitemShut {NoStop}%
\bibitem [{\citenamefont {Salvati}\ \emph {et~al.}(2023)\citenamefont
  {Salvati}, \citenamefont {Salmon}, \citenamefont {Barnes},\ and\
  \citenamefont {Arvidsson-Shukur}}]{Salvati23}%
  \BibitemOpen
  \bibfield  {author} {\bibinfo {author} {\bibfnamefont {F.}~\bibnamefont
  {Salvati}}, \bibinfo {author} {\bibfnamefont {W.}~\bibnamefont {Salmon}},
  \bibinfo {author} {\bibfnamefont {C.~H.~W.}\ \bibnamefont {Barnes}}, \ and\
  \bibinfo {author} {\bibfnamefont {D.~R.~M.}\ \bibnamefont
  {Arvidsson-Shukur}},\ }\href {https://arxiv.org/abs/2307.08648} {\  (\bibinfo
  {year} {2023})},\ \Eprint {http://arxiv.org/abs/2307.08648} {arXiv:2307.08648
  [quant-ph]} \BibitemShut {NoStop}%
\bibitem [{\citenamefont {Aharonov}\ \emph {et~al.}(1988)\citenamefont
  {Aharonov}, \citenamefont {Albert},\ and\ \citenamefont
  {Vaidman}}]{Vaidman88}%
  \BibitemOpen
  \bibfield  {author} {\bibinfo {author} {\bibfnamefont {Y.}~\bibnamefont
  {Aharonov}}, \bibinfo {author} {\bibfnamefont {D.~Z.}\ \bibnamefont
  {Albert}}, \ and\ \bibinfo {author} {\bibfnamefont {L.}~\bibnamefont
  {Vaidman}},\ }\href {\doibase 10.1103/PhysRevLett.60.1351} {\bibfield
  {journal} {\bibinfo  {journal} {Phys. Rev. Lett.}\ }\textbf {\bibinfo
  {volume} {60}},\ \bibinfo {pages} {1351} (\bibinfo {year}
  {1988})}\BibitemShut {NoStop}%
\bibitem [{\citenamefont {Duck}\ \emph {et~al.}(1989)\citenamefont {Duck},
  \citenamefont {Stevenson},\ and\ \citenamefont {Sudarshan}}]{Duck89}%
  \BibitemOpen
  \bibfield  {author} {\bibinfo {author} {\bibfnamefont {I.~M.}\ \bibnamefont
  {Duck}}, \bibinfo {author} {\bibfnamefont {P.~M.}\ \bibnamefont {Stevenson}},
  \ and\ \bibinfo {author} {\bibfnamefont {E.~C.~G.}\ \bibnamefont
  {Sudarshan}},\ }\href {\doibase 10.1103/PhysRevD.40.2112} {\bibfield
  {journal} {\bibinfo  {journal} {Phys. Rev. D}\ }\textbf {\bibinfo {volume}
  {40}},\ \bibinfo {pages} {2112} (\bibinfo {year} {1989})}\BibitemShut
  {NoStop}%
\bibitem [{\citenamefont {Hosten}\ and\ \citenamefont
  {Kwiat}(2008)}]{Hosten08}%
  \BibitemOpen
  \bibfield  {author} {\bibinfo {author} {\bibfnamefont {O.}~\bibnamefont
  {Hosten}}\ and\ \bibinfo {author} {\bibfnamefont {P.}~\bibnamefont {Kwiat}},\
  }\href {\doibase 10.1126/science.1152697} {\bibfield  {journal} {\bibinfo
  {journal} {Science}\ }\textbf {\bibinfo {volume} {319}},\ \bibinfo {pages}
  {787} (\bibinfo {year} {2008})}\BibitemShut {NoStop}%
\bibitem [{\citenamefont {Tollaksen}(2007)}]{Tollaksen07}%
  \BibitemOpen
  \bibfield  {author} {\bibinfo {author} {\bibfnamefont {J.}~\bibnamefont
  {Tollaksen}},\ }\href@noop {} {\bibfield  {journal} {\bibinfo  {journal} {J.
  Phys. A}\ }\textbf {\bibinfo {volume} {40}},\ \bibinfo {pages} {9033}
  (\bibinfo {year} {2007})}\BibitemShut {NoStop}%
\bibitem [{\citenamefont {Pusey}(2014)}]{Pusey14}%
  \BibitemOpen
  \bibfield  {author} {\bibinfo {author} {\bibfnamefont {M.~F.}\ \bibnamefont
  {Pusey}},\ }\href {\doibase 10.1103/PhysRevLett.113.200401} {\bibfield
  {journal} {\bibinfo  {journal} {Phys. Rev. Lett.}\ }\textbf {\bibinfo
  {volume} {113}},\ \bibinfo {pages} {200401} (\bibinfo {year}
  {2014})}\BibitemShut {NoStop}%
\bibitem [{\citenamefont {Kunjwal}\ \emph {et~al.}(2019)\citenamefont
  {Kunjwal}, \citenamefont {Lostaglio},\ and\ \citenamefont
  {Pusey}}]{Kunjwal19}%
  \BibitemOpen
  \bibfield  {author} {\bibinfo {author} {\bibfnamefont {R.}~\bibnamefont
  {Kunjwal}}, \bibinfo {author} {\bibfnamefont {M.}~\bibnamefont {Lostaglio}},
  \ and\ \bibinfo {author} {\bibfnamefont {M.~F.}\ \bibnamefont {Pusey}},\
  }\href {\doibase 10.1103/PhysRevA.100.042116} {\bibfield  {journal} {\bibinfo
   {journal} {Phys. Rev. A}\ }\textbf {\bibinfo {volume} {100}},\ \bibinfo
  {pages} {042116} (\bibinfo {year} {2019})}\BibitemShut {NoStop}%
\bibitem [{\citenamefont {Aharonov}\ and\ \citenamefont
  {Vaidman}(2008)}]{Aharonov08}%
  \BibitemOpen
  \bibfield  {author} {\bibinfo {author} {\bibfnamefont {Y.}~\bibnamefont
  {Aharonov}}\ and\ \bibinfo {author} {\bibfnamefont {L.}~\bibnamefont
  {Vaidman}},\ }\href@noop {} {\emph {\bibinfo {title} {Time in quantum
  mechanics}}}\ (\bibinfo  {publisher} {Springer},\ \bibinfo {year} {2008})\
  pp.\ \bibinfo {pages} {399--447}\BibitemShut {NoStop}%
\bibitem [{\citenamefont {Leifer}\ and\ \citenamefont
  {Pusey}(2017)}]{Leifer17}%
  \BibitemOpen
  \bibfield  {author} {\bibinfo {author} {\bibfnamefont {M.~S.}\ \bibnamefont
  {Leifer}}\ and\ \bibinfo {author} {\bibfnamefont {M.~F.}\ \bibnamefont
  {Pusey}},\ }\href@noop {} {\bibfield  {journal} {\bibinfo  {journal}
  {Proceedings of the Royal Society A: Mathematical, Physical and Engineering
  Sciences}\ }\textbf {\bibinfo {volume} {473}},\ \bibinfo {pages} {20160607}
  (\bibinfo {year} {2017})}\BibitemShut {NoStop}%
\bibitem [{\citenamefont {G\"odel}(1949)}]{Godel49}%
  \BibitemOpen
  \bibfield  {author} {\bibinfo {author} {\bibfnamefont {K.}~\bibnamefont
  {G\"odel}},\ }\href {\doibase 10.1103/RevModPhys.21.447} {\bibfield
  {journal} {\bibinfo  {journal} {Rev. Mod. Phys.}\ }\textbf {\bibinfo {volume}
  {21}},\ \bibinfo {pages} {447} (\bibinfo {year} {1949})}\BibitemShut
  {NoStop}%
\bibitem [{\citenamefont {Morris}\ \emph {et~al.}(1988)\citenamefont {Morris},
  \citenamefont {Thorne},\ and\ \citenamefont {Yurtsever}}]{Morris88}%
  \BibitemOpen
  \bibfield  {author} {\bibinfo {author} {\bibfnamefont {M.~S.}\ \bibnamefont
  {Morris}}, \bibinfo {author} {\bibfnamefont {K.~S.}\ \bibnamefont {Thorne}},
  \ and\ \bibinfo {author} {\bibfnamefont {U.}~\bibnamefont {Yurtsever}},\
  }\href {\doibase 10.1103/PhysRevLett.61.1446} {\bibfield  {journal} {\bibinfo
   {journal} {Phys. Rev. Lett.}\ }\textbf {\bibinfo {volume} {61}},\ \bibinfo
  {pages} {1446} (\bibinfo {year} {1988})}\BibitemShut {NoStop}%
\bibitem [{\citenamefont {Deutsch}(1991)}]{Deutsch91}%
  \BibitemOpen
  \bibfield  {author} {\bibinfo {author} {\bibfnamefont {D.}~\bibnamefont
  {Deutsch}},\ }\href@noop {} {\bibfield  {journal} {\bibinfo  {journal}
  {Physical Review D}\ }\textbf {\bibinfo {volume} {44}},\ \bibinfo {pages}
  {3197} (\bibinfo {year} {1991})}\BibitemShut {NoStop}%
\bibitem [{\citenamefont {Bennett}(2005)}]{Bennett05}%
  \BibitemOpen
  \bibfield  {author} {\bibinfo {author} {\bibfnamefont {C.~H.}\ \bibnamefont
  {Bennett}},\ }in\ \href@noop {} {\emph {\bibinfo {booktitle} {Proceedings of
  QUPON}}}\ (\bibinfo {address} {Wien},\ \bibinfo {year} {2005})\BibitemShut
  {NoStop}%
\bibitem [{\citenamefont {Svetlichny}(2011)}]{Svetlichny11}%
  \BibitemOpen
  \bibfield  {author} {\bibinfo {author} {\bibfnamefont {G.}~\bibnamefont
  {Svetlichny}},\ }\href {\doibase 10.1007/s10773-011-0973-x} {\bibfield
  {journal} {\bibinfo  {journal} {International Journal of Theoretical
  Physics}\ }\textbf {\bibinfo {volume} {50}},\ \bibinfo {pages} {3903}
  (\bibinfo {year} {2011})}\BibitemShut {NoStop}%
\bibitem [{\citenamefont {Lloyd}\ \emph
  {et~al.}(2011{\natexlab{a}})\citenamefont {Lloyd}, \citenamefont {Maccone},
  \citenamefont {Garcia-Patron}, \citenamefont {Giovannetti}, \citenamefont
  {Shikano}, \citenamefont {Pirandola}, \citenamefont {Rozema}, \citenamefont
  {Darabi}, \citenamefont {Soudagar}, \citenamefont {Shalm},\ and\
  \citenamefont {Steinberg}}]{Lloyd11}%
  \BibitemOpen
  \bibfield  {author} {\bibinfo {author} {\bibfnamefont {S.}~\bibnamefont
  {Lloyd}}, \bibinfo {author} {\bibfnamefont {L.}~\bibnamefont {Maccone}},
  \bibinfo {author} {\bibfnamefont {R.}~\bibnamefont {Garcia-Patron}}, \bibinfo
  {author} {\bibfnamefont {V.}~\bibnamefont {Giovannetti}}, \bibinfo {author}
  {\bibfnamefont {Y.}~\bibnamefont {Shikano}}, \bibinfo {author} {\bibfnamefont
  {S.}~\bibnamefont {Pirandola}}, \bibinfo {author} {\bibfnamefont {L.~A.}\
  \bibnamefont {Rozema}}, \bibinfo {author} {\bibfnamefont {A.}~\bibnamefont
  {Darabi}}, \bibinfo {author} {\bibfnamefont {Y.}~\bibnamefont {Soudagar}},
  \bibinfo {author} {\bibfnamefont {L.~K.}\ \bibnamefont {Shalm}}, \ and\
  \bibinfo {author} {\bibfnamefont {A.~M.}\ \bibnamefont {Steinberg}},\ }\href
  {\doibase 10.1103/PhysRevLett.106.040403} {\bibfield  {journal} {\bibinfo
  {journal} {Phys. Rev. Lett.}\ }\textbf {\bibinfo {volume} {106}},\ \bibinfo
  {pages} {040403} (\bibinfo {year} {2011}{\natexlab{a}})}\BibitemShut
  {NoStop}%
\bibitem [{\citenamefont {Lloyd}\ \emph
  {et~al.}(2011{\natexlab{b}})\citenamefont {Lloyd}, \citenamefont {Maccone},
  \citenamefont {Garcia-Patron}, \citenamefont {Giovannetti},\ and\
  \citenamefont {Shikano}}]{Lloyd11-2}%
  \BibitemOpen
  \bibfield  {author} {\bibinfo {author} {\bibfnamefont {S.}~\bibnamefont
  {Lloyd}}, \bibinfo {author} {\bibfnamefont {L.}~\bibnamefont {Maccone}},
  \bibinfo {author} {\bibfnamefont {R.}~\bibnamefont {Garcia-Patron}}, \bibinfo
  {author} {\bibfnamefont {V.}~\bibnamefont {Giovannetti}}, \ and\ \bibinfo
  {author} {\bibfnamefont {Y.}~\bibnamefont {Shikano}},\ }\href {\doibase
  10.1103/PhysRevD.84.025007} {\bibfield  {journal} {\bibinfo  {journal} {Phys.
  Rev. D}\ }\textbf {\bibinfo {volume} {84}},\ \bibinfo {pages} {025007}
  (\bibinfo {year} {2011}{\natexlab{b}})}\BibitemShut {NoStop}%
\bibitem [{\citenamefont {Gott}(1991)}]{Gott91}%
  \BibitemOpen
  \bibfield  {author} {\bibinfo {author} {\bibfnamefont {J.~R.}\ \bibnamefont
  {Gott}},\ }\href {\doibase 10.1103/PhysRevLett.66.1126} {\bibfield  {journal}
  {\bibinfo  {journal} {Phys. Rev. Lett.}\ }\textbf {\bibinfo {volume} {66}},\
  \bibinfo {pages} {1126} (\bibinfo {year} {1991})}\BibitemShut {NoStop}%
\bibitem [{\citenamefont {Hawking}(1992)}]{Hawking92}%
  \BibitemOpen
  \bibfield  {author} {\bibinfo {author} {\bibfnamefont {S.~W.}\ \bibnamefont
  {Hawking}},\ }\href {\doibase 10.1103/PhysRevD.46.603} {\bibfield  {journal}
  {\bibinfo  {journal} {Phys. Rev. D}\ }\textbf {\bibinfo {volume} {46}},\
  \bibinfo {pages} {603} (\bibinfo {year} {1992})}\BibitemShut {NoStop}%
\bibitem [{\citenamefont {Deser}\ \emph {et~al.}(1992)\citenamefont {Deser},
  \citenamefont {Jackiw},\ and\ \citenamefont {'t~Hooft}}]{Deser92}%
  \BibitemOpen
  \bibfield  {author} {\bibinfo {author} {\bibfnamefont {S.}~\bibnamefont
  {Deser}}, \bibinfo {author} {\bibfnamefont {R.}~\bibnamefont {Jackiw}}, \
  and\ \bibinfo {author} {\bibfnamefont {G.}~\bibnamefont {'t~Hooft}},\ }\href
  {\doibase 10.1103/PhysRevLett.68.267} {\bibfield  {journal} {\bibinfo
  {journal} {Phys. Rev. Lett.}\ }\textbf {\bibinfo {volume} {68}},\ \bibinfo
  {pages} {267} (\bibinfo {year} {1992})}\BibitemShut {NoStop}%
\bibitem [{\citenamefont {Carroll}\ \emph {et~al.}(1994)\citenamefont
  {Carroll}, \citenamefont {Farhi}, \citenamefont {Guth},\ and\ \citenamefont
  {Olum}}]{Carroll94}%
  \BibitemOpen
  \bibfield  {author} {\bibinfo {author} {\bibfnamefont {S.~M.}\ \bibnamefont
  {Carroll}}, \bibinfo {author} {\bibfnamefont {E.}~\bibnamefont {Farhi}},
  \bibinfo {author} {\bibfnamefont {A.~H.}\ \bibnamefont {Guth}}, \ and\
  \bibinfo {author} {\bibfnamefont {K.~D.}\ \bibnamefont {Olum}},\ }\href
  {\doibase 10.1103/PhysRevD.50.6190} {\bibfield  {journal} {\bibinfo
  {journal} {Phys. Rev. D}\ }\textbf {\bibinfo {volume} {50}},\ \bibinfo
  {pages} {6190} (\bibinfo {year} {1994})}\BibitemShut {NoStop}%
\bibitem [{\citenamefont {Allen}(2014)}]{Allen14}%
  \BibitemOpen
  \bibfield  {author} {\bibinfo {author} {\bibfnamefont {J.-M.~A.}\
  \bibnamefont {Allen}},\ }\href {\doibase 10.1103/PhysRevA.90.042107}
  {\bibfield  {journal} {\bibinfo  {journal} {Phys. Rev. A}\ }\textbf {\bibinfo
  {volume} {90}},\ \bibinfo {pages} {042107} (\bibinfo {year}
  {2014})}\BibitemShut {NoStop}%
\bibitem [{\citenamefont {Brun}\ and\ \citenamefont {Wilde}(2017)}]{Brun17}%
  \BibitemOpen
  \bibfield  {author} {\bibinfo {author} {\bibfnamefont {T.~A.}\ \bibnamefont
  {Brun}}\ and\ \bibinfo {author} {\bibfnamefont {M.~M.}\ \bibnamefont
  {Wilde}},\ }\href {\doibase 10.1007/s10701-017-0066-7} {\bibfield  {journal}
  {\bibinfo  {journal} {Foundations of Physics}\ }\textbf {\bibinfo {volume}
  {47}},\ \bibinfo {pages} {375} (\bibinfo {year} {2017})}\BibitemShut
  {NoStop}%
\bibitem [{\citenamefont {Nielsen}\ and\ \citenamefont
  {Chuang}(2011)}]{Nielsen11}%
  \BibitemOpen
  \bibfield  {author} {\bibinfo {author} {\bibfnamefont {M.~A.}\ \bibnamefont
  {Nielsen}}\ and\ \bibinfo {author} {\bibfnamefont {I.~L.}\ \bibnamefont
  {Chuang}},\ }\href@noop {} {\emph {\bibinfo {title} {Quantum Computation and
  Quantum Information: 10th Anniversary Edition}}},\ \bibinfo {edition} {10th}\
  ed.\ (\bibinfo  {publisher} {Cambridge University Press},\ \bibinfo {address}
  {New York, NY, USA},\ \bibinfo {year} {2011})\BibitemShut {NoStop}%
\bibitem [{\citenamefont {Bennett}\ \emph {et~al.}(1993)\citenamefont
  {Bennett}, \citenamefont {Brassard}, \citenamefont {Cr\'epeau}, \citenamefont
  {Jozsa}, \citenamefont {Peres},\ and\ \citenamefont {Wootters}}]{Bennett93}%
  \BibitemOpen
  \bibfield  {author} {\bibinfo {author} {\bibfnamefont {C.~H.}\ \bibnamefont
  {Bennett}}, \bibinfo {author} {\bibfnamefont {G.}~\bibnamefont {Brassard}},
  \bibinfo {author} {\bibfnamefont {C.}~\bibnamefont {Cr\'epeau}}, \bibinfo
  {author} {\bibfnamefont {R.}~\bibnamefont {Jozsa}}, \bibinfo {author}
  {\bibfnamefont {A.}~\bibnamefont {Peres}}, \ and\ \bibinfo {author}
  {\bibfnamefont {W.~K.}\ \bibnamefont {Wootters}},\ }\href {\doibase
  10.1103/PhysRevLett.70.1895} {\bibfield  {journal} {\bibinfo  {journal}
  {Phys. Rev. Lett.}\ }\textbf {\bibinfo {volume} {70}},\ \bibinfo {pages}
  {1895} (\bibinfo {year} {1993})}\BibitemShut {NoStop}%
\bibitem [{\citenamefont {Coecke}(2004)}]{Coecke04}%
  \BibitemOpen
  \bibfield  {author} {\bibinfo {author} {\bibfnamefont {B.}~\bibnamefont
  {Coecke}},\ }\href {https://arxiv.org/abs/quant-ph/0402014} {\  (\bibinfo
  {year} {2004})}\BibitemShut {NoStop}%
\bibitem [{\citenamefont {Braunstein}\ and\ \citenamefont
  {Caves}(1994)}]{Braunstein94}%
  \BibitemOpen
  \bibfield  {author} {\bibinfo {author} {\bibfnamefont {S.~L.}\ \bibnamefont
  {Braunstein}}\ and\ \bibinfo {author} {\bibfnamefont {C.~M.}\ \bibnamefont
  {Caves}},\ }\href {\doibase 10.1103/PhysRevLett.72.3439} {\bibfield
  {journal} {\bibinfo  {journal} {Phys. Rev. Lett.}\ }\textbf {\bibinfo
  {volume} {72}},\ \bibinfo {pages} {3439} (\bibinfo {year}
  {1994})}\BibitemShut {NoStop}%
\bibitem [{\citenamefont {Combes}\ \emph {et~al.}(2014)\citenamefont {Combes},
  \citenamefont {Ferrie}, \citenamefont {Jiang},\ and\ \citenamefont
  {Caves}}]{Combes14}%
  \BibitemOpen
  \bibfield  {author} {\bibinfo {author} {\bibfnamefont {J.}~\bibnamefont
  {Combes}}, \bibinfo {author} {\bibfnamefont {C.}~\bibnamefont {Ferrie}},
  \bibinfo {author} {\bibfnamefont {Z.}~\bibnamefont {Jiang}}, \ and\ \bibinfo
  {author} {\bibfnamefont {C.~M.}\ \bibnamefont {Caves}},\ }\href {\doibase
  10.1103/PhysRevA.89.052117} {\bibfield  {journal} {\bibinfo  {journal} {Phys.
  Rev. A}\ }\textbf {\bibinfo {volume} {89}},\ \bibinfo {pages} {052117}
  (\bibinfo {year} {2014})}\BibitemShut {NoStop}%
\bibitem [{\citenamefont {Ferrie}\ and\ \citenamefont
  {Combes}(2014)}]{Ferrie14-2}%
  \BibitemOpen
  \bibfield  {author} {\bibinfo {author} {\bibfnamefont {C.}~\bibnamefont
  {Ferrie}}\ and\ \bibinfo {author} {\bibfnamefont {J.}~\bibnamefont
  {Combes}},\ }\href {\doibase 10.1103/PhysRevLett.112.040406} {\bibfield
  {journal} {\bibinfo  {journal} {Phys. Rev. Lett.}\ }\textbf {\bibinfo
  {volume} {112}},\ \bibinfo {pages} {040406} (\bibinfo {year}
  {2014})}\BibitemShut {NoStop}%
\bibitem [{\citenamefont {Arvidsson-Shukur}\ \emph {et~al.}(2021)\citenamefont
  {Arvidsson-Shukur}, \citenamefont {Drori},\ and\ \citenamefont
  {Halpern}}]{ArvShuk21-2}%
  \BibitemOpen
  \bibfield  {author} {\bibinfo {author} {\bibfnamefont {D.~R.~M.}\
  \bibnamefont {Arvidsson-Shukur}}, \bibinfo {author} {\bibfnamefont {J.~C.}\
  \bibnamefont {Drori}}, \ and\ \bibinfo {author} {\bibfnamefont {N.~Y.}\
  \bibnamefont {Halpern}},\ }\href {\doibase 10.1088/1751-8121/ac0289}
  {\bibfield  {journal} {\bibinfo  {journal} {Journal of Physics A:
  Mathematical and Theoretical}\ }\textbf {\bibinfo {volume} {54}},\ \bibinfo
  {pages} {284001} (\bibinfo {year} {2021})}\BibitemShut {NoStop}%
\bibitem [{\citenamefont {Spekkens}(2005)}]{Spekkens05}%
  \BibitemOpen
  \bibfield  {author} {\bibinfo {author} {\bibfnamefont {R.~W.}\ \bibnamefont
  {Spekkens}},\ }\href {\doibase 10.1103/PhysRevA.71.052108} {\bibfield
  {journal} {\bibinfo  {journal} {Phys. Rev. A}\ }\textbf {\bibinfo {volume}
  {71}},\ \bibinfo {pages} {052108} (\bibinfo {year} {2005})}\BibitemShut
  {NoStop}%
\bibitem [{\citenamefont {Lostaglio}(2020)}]{Lostaglio20}%
  \BibitemOpen
  \bibfield  {author} {\bibinfo {author} {\bibfnamefont {M.}~\bibnamefont
  {Lostaglio}},\ }\href {https://arxiv.org/abs/2004.01213} {\bibfield
  {journal} {\bibinfo  {journal} {arXiv preprint arXiv:2004.01213}\ } (\bibinfo
  {year} {2020})}\BibitemShut {NoStop}%
\bibitem [{\citenamefont {Giovannetti}\ \emph {et~al.}(2006)\citenamefont
  {Giovannetti}, \citenamefont {Lloyd},\ and\ \citenamefont
  {Maccone}}]{Giovanetti06}%
  \BibitemOpen
  \bibfield  {author} {\bibinfo {author} {\bibfnamefont {V.}~\bibnamefont
  {Giovannetti}}, \bibinfo {author} {\bibfnamefont {S.}~\bibnamefont {Lloyd}},
  \ and\ \bibinfo {author} {\bibfnamefont {L.}~\bibnamefont {Maccone}},\
  }\href@noop {} {\bibfield  {journal} {\bibinfo  {journal} {Physical review
  letters}\ }\textbf {\bibinfo {volume} {96}},\ \bibinfo {pages} {010401}
  (\bibinfo {year} {2006})}\BibitemShut {NoStop}%
\bibitem [{\citenamefont {Barbieri}\ \emph {et~al.}(2005)\citenamefont
  {Barbieri}, \citenamefont {Cinelli}, \citenamefont {Mataloni},\ and\
  \citenamefont {De~Martini}}]{Barbieri05}%
  \BibitemOpen
  \bibfield  {author} {\bibinfo {author} {\bibfnamefont {M.}~\bibnamefont
  {Barbieri}}, \bibinfo {author} {\bibfnamefont {C.}~\bibnamefont {Cinelli}},
  \bibinfo {author} {\bibfnamefont {P.}~\bibnamefont {Mataloni}}, \ and\
  \bibinfo {author} {\bibfnamefont {F.}~\bibnamefont {De~Martini}},\ }\href
  {\doibase 10.1103/PhysRevA.72.052110} {\bibfield  {journal} {\bibinfo
  {journal} {Phys. Rev. A}\ }\textbf {\bibinfo {volume} {72}},\ \bibinfo
  {pages} {052110} (\bibinfo {year} {2005})}\BibitemShut {NoStop}%
\bibitem [{\citenamefont {Cinelli}\ \emph {et~al.}(2005)\citenamefont
  {Cinelli}, \citenamefont {Barbieri}, \citenamefont {Perris}, \citenamefont
  {Mataloni},\ and\ \citenamefont {Martini}}]{Cinelli05}%
  \BibitemOpen
  \bibfield  {author} {\bibinfo {author} {\bibfnamefont {C.}~\bibnamefont
  {Cinelli}}, \bibinfo {author} {\bibfnamefont {M.}~\bibnamefont {Barbieri}},
  \bibinfo {author} {\bibfnamefont {R.}~\bibnamefont {Perris}}, \bibinfo
  {author} {\bibfnamefont {P.}~\bibnamefont {Mataloni}}, \ and\ \bibinfo
  {author} {\bibfnamefont {F.~D.}\ \bibnamefont {Martini}},\ }\href {\doibase
  10.1103/PhysRevLett.95.240405} {\bibfield  {journal} {\bibinfo  {journal}
  {Phys. Rev. Lett.}\ }\textbf {\bibinfo {volume} {95}},\ \bibinfo {pages}
  {240405} (\bibinfo {year} {2005})}\BibitemShut {NoStop}%
\bibitem [{\citenamefont {Barreiro}\ \emph {et~al.}(2005)\citenamefont
  {Barreiro}, \citenamefont {Langford}, \citenamefont {Peters},\ and\
  \citenamefont {Kwiat}}]{Barreiro05}%
  \BibitemOpen
  \bibfield  {author} {\bibinfo {author} {\bibfnamefont {J.~T.}\ \bibnamefont
  {Barreiro}}, \bibinfo {author} {\bibfnamefont {N.~K.}\ \bibnamefont
  {Langford}}, \bibinfo {author} {\bibfnamefont {N.~A.}\ \bibnamefont
  {Peters}}, \ and\ \bibinfo {author} {\bibfnamefont {P.~G.}\ \bibnamefont
  {Kwiat}},\ }\href {\doibase 10.1103/PhysRevLett.95.260501} {\bibfield
  {journal} {\bibinfo  {journal} {Phys. Rev. Lett.}\ }\textbf {\bibinfo
  {volume} {95}},\ \bibinfo {pages} {260501} (\bibinfo {year}
  {2005})}\BibitemShut {NoStop}%
\bibitem [{\citenamefont {Vallone}\ \emph {et~al.}(2007)\citenamefont
  {Vallone}, \citenamefont {Pomarico}, \citenamefont {Mataloni}, \citenamefont
  {De~Martini},\ and\ \citenamefont {Berardi}}]{Vallone07}%
  \BibitemOpen
  \bibfield  {author} {\bibinfo {author} {\bibfnamefont {G.}~\bibnamefont
  {Vallone}}, \bibinfo {author} {\bibfnamefont {E.}~\bibnamefont {Pomarico}},
  \bibinfo {author} {\bibfnamefont {P.}~\bibnamefont {Mataloni}}, \bibinfo
  {author} {\bibfnamefont {F.}~\bibnamefont {De~Martini}}, \ and\ \bibinfo
  {author} {\bibfnamefont {V.}~\bibnamefont {Berardi}},\ }\href {\doibase
  10.1103/PhysRevLett.98.180502} {\bibfield  {journal} {\bibinfo  {journal}
  {Phys. Rev. Lett.}\ }\textbf {\bibinfo {volume} {98}},\ \bibinfo {pages}
  {180502} (\bibinfo {year} {2007})}\BibitemShut {NoStop}%
\bibitem [{\citenamefont {Chen}\ \emph {et~al.}(2007)\citenamefont {Chen},
  \citenamefont {Li}, \citenamefont {Zhang}, \citenamefont {Chen},
  \citenamefont {Goebel}, \citenamefont {Chen}, \citenamefont {Mair},\ and\
  \citenamefont {Pan}}]{Chen07}%
  \BibitemOpen
  \bibfield  {author} {\bibinfo {author} {\bibfnamefont {K.}~\bibnamefont
  {Chen}}, \bibinfo {author} {\bibfnamefont {C.-M.}\ \bibnamefont {Li}},
  \bibinfo {author} {\bibfnamefont {Q.}~\bibnamefont {Zhang}}, \bibinfo
  {author} {\bibfnamefont {Y.-A.}\ \bibnamefont {Chen}}, \bibinfo {author}
  {\bibfnamefont {A.}~\bibnamefont {Goebel}}, \bibinfo {author} {\bibfnamefont
  {S.}~\bibnamefont {Chen}}, \bibinfo {author} {\bibfnamefont {A.}~\bibnamefont
  {Mair}}, \ and\ \bibinfo {author} {\bibfnamefont {J.-W.}\ \bibnamefont
  {Pan}},\ }\href {\doibase 10.1103/PhysRevLett.99.120503} {\bibfield
  {journal} {\bibinfo  {journal} {Phys. Rev. Lett.}\ }\textbf {\bibinfo
  {volume} {99}},\ \bibinfo {pages} {120503} (\bibinfo {year}
  {2007})}\BibitemShut {NoStop}%
\bibitem [{\citenamefont {Ciampini}\ \emph {et~al.}(2016)\citenamefont
  {Ciampini}, \citenamefont {Orieux}, \citenamefont {Paesani}, \citenamefont
  {Sciarrino}, \citenamefont {Corrielli}, \citenamefont {Crespi}, \citenamefont
  {Ramponi}, \citenamefont {Osellame},\ and\ \citenamefont
  {Mataloni}}]{Ciampini16}%
  \BibitemOpen
  \bibfield  {author} {\bibinfo {author} {\bibfnamefont {M.~A.}\ \bibnamefont
  {Ciampini}}, \bibinfo {author} {\bibfnamefont {A.}~\bibnamefont {Orieux}},
  \bibinfo {author} {\bibfnamefont {S.}~\bibnamefont {Paesani}}, \bibinfo
  {author} {\bibfnamefont {F.}~\bibnamefont {Sciarrino}}, \bibinfo {author}
  {\bibfnamefont {G.}~\bibnamefont {Corrielli}}, \bibinfo {author}
  {\bibfnamefont {A.}~\bibnamefont {Crespi}}, \bibinfo {author} {\bibfnamefont
  {R.}~\bibnamefont {Ramponi}}, \bibinfo {author} {\bibfnamefont
  {R.}~\bibnamefont {Osellame}}, \ and\ \bibinfo {author} {\bibfnamefont
  {P.}~\bibnamefont {Mataloni}},\ }\href@noop {} {\bibfield  {journal}
  {\bibinfo  {journal} {Light: Science \& Applications}\ }\textbf {\bibinfo
  {volume} {5}},\ \bibinfo {pages} {e16064} (\bibinfo {year}
  {2016})}\BibitemShut {NoStop}%
\bibitem [{\citenamefont {Brun}(2003)}]{Brun03}%
  \BibitemOpen
  \bibfield  {author} {\bibinfo {author} {\bibfnamefont {T.~A.}\ \bibnamefont
  {Brun}},\ }\href {\doibase 10.1023/A:1025967225931} {\bibfield  {journal}
  {\bibinfo  {journal} {Foundations of Physics Letters}\ }\textbf {\bibinfo
  {volume} {16}},\ \bibinfo {pages} {245} (\bibinfo {year} {2003})}\BibitemShut
  {NoStop}%
\bibitem [{\citenamefont {Aaronson}(2004)}]{Aaronson04}%
  \BibitemOpen
  \bibfield  {author} {\bibinfo {author} {\bibfnamefont {S.}~\bibnamefont
  {Aaronson}},\ }\href {\doibase 10.48550/ARXIV.QUANT-PH/0412187} {\  (\bibinfo
  {year} {2004}),\ 10.48550/ARXIV.QUANT-PH/0412187}\BibitemShut {NoStop}%
\bibitem [{\citenamefont {Aaronson}\ and\ \citenamefont
  {Watrous}(2009)}]{Aaronson09}%
  \BibitemOpen
  \bibfield  {author} {\bibinfo {author} {\bibfnamefont {S.}~\bibnamefont
  {Aaronson}}\ and\ \bibinfo {author} {\bibfnamefont {J.}~\bibnamefont
  {Watrous}},\ }\href@noop {} {\bibfield  {journal} {\bibinfo  {journal}
  {Proceedings of the Royal Society A: Mathematical, Physical and Engineering
  Sciences}\ }\textbf {\bibinfo {volume} {465}},\ \bibinfo {pages} {631}
  (\bibinfo {year} {2009})}\BibitemShut {NoStop}%
\bibitem [{\citenamefont {Brun}\ \emph {et~al.}(2009)\citenamefont {Brun},
  \citenamefont {Harrington},\ and\ \citenamefont {Wilde}}]{Brun09}%
  \BibitemOpen
  \bibfield  {author} {\bibinfo {author} {\bibfnamefont {T.~A.}\ \bibnamefont
  {Brun}}, \bibinfo {author} {\bibfnamefont {J.}~\bibnamefont {Harrington}}, \
  and\ \bibinfo {author} {\bibfnamefont {M.~M.}\ \bibnamefont {Wilde}},\ }\href
  {\doibase 10.1103/PhysRevLett.102.210402} {\bibfield  {journal} {\bibinfo
  {journal} {Phys. Rev. Lett.}\ }\textbf {\bibinfo {volume} {102}},\ \bibinfo
  {pages} {210402} (\bibinfo {year} {2009})}\BibitemShut {NoStop}%
\bibitem [{\citenamefont {Brun}\ and\ \citenamefont {Wilde}(2012)}]{Brun12}%
  \BibitemOpen
  \bibfield  {author} {\bibinfo {author} {\bibfnamefont {T.~A.}\ \bibnamefont
  {Brun}}\ and\ \bibinfo {author} {\bibfnamefont {M.~M.}\ \bibnamefont
  {Wilde}},\ }\href {\doibase 10.1007/s10701-011-9601-0} {\bibfield  {journal}
  {\bibinfo  {journal} {Foundations of Physics}\ }\textbf {\bibinfo {volume}
  {42}},\ \bibinfo {pages} {341} (\bibinfo {year} {2012})}\BibitemShut
  {NoStop}%
\bibitem [{\citenamefont {Brun}\ \emph {et~al.}(2013)\citenamefont {Brun},
  \citenamefont {Wilde},\ and\ \citenamefont {Winter}}]{Brun13}%
  \BibitemOpen
  \bibfield  {author} {\bibinfo {author} {\bibfnamefont {T.~A.}\ \bibnamefont
  {Brun}}, \bibinfo {author} {\bibfnamefont {M.~M.}\ \bibnamefont {Wilde}}, \
  and\ \bibinfo {author} {\bibfnamefont {A.}~\bibnamefont {Winter}},\ }\href
  {\doibase 10.1103/PhysRevLett.111.190401} {\bibfield  {journal} {\bibinfo
  {journal} {Phys. Rev. Lett.}\ }\textbf {\bibinfo {volume} {111}},\ \bibinfo
  {pages} {190401} (\bibinfo {year} {2013})}\BibitemShut {NoStop}%
\bibitem [{\citenamefont {Pienaar}\ \emph {et~al.}(2013)\citenamefont
  {Pienaar}, \citenamefont {Ralph},\ and\ \citenamefont {Myers}}]{Pienaar13}%
  \BibitemOpen
  \bibfield  {author} {\bibinfo {author} {\bibfnamefont {J.~L.}\ \bibnamefont
  {Pienaar}}, \bibinfo {author} {\bibfnamefont {T.~C.}\ \bibnamefont {Ralph}},
  \ and\ \bibinfo {author} {\bibfnamefont {C.~R.}\ \bibnamefont {Myers}},\
  }\href {\doibase 10.1103/PhysRevLett.110.060501} {\bibfield  {journal}
  {\bibinfo  {journal} {Phys. Rev. Lett.}\ }\textbf {\bibinfo {volume} {110}},\
  \bibinfo {pages} {060501} (\bibinfo {year} {2013})}\BibitemShut {NoStop}%
\bibitem [{\citenamefont {Bub}\ and\ \citenamefont {Stairs}(2014)}]{Bub14}%
  \BibitemOpen
  \bibfield  {author} {\bibinfo {author} {\bibfnamefont {J.}~\bibnamefont
  {Bub}}\ and\ \bibinfo {author} {\bibfnamefont {A.}~\bibnamefont {Stairs}},\
  }\href {\doibase 10.1103/PhysRevA.89.022311} {\bibfield  {journal} {\bibinfo
  {journal} {Phys. Rev. A}\ }\textbf {\bibinfo {volume} {89}},\ \bibinfo
  {pages} {022311} (\bibinfo {year} {2014})}\BibitemShut {NoStop}%
\bibitem [{\citenamefont {Bartkiewicz}\ \emph {et~al.}(2019)\citenamefont
  {Bartkiewicz}, \citenamefont {Grudka}, \citenamefont {Horodecki},
  \citenamefont {\L{}odyga},\ and\ \citenamefont
  {Wychowaniec}}]{Bartkiewicz19}%
  \BibitemOpen
  \bibfield  {author} {\bibinfo {author} {\bibfnamefont {M.}~\bibnamefont
  {Bartkiewicz}}, \bibinfo {author} {\bibfnamefont {A.}~\bibnamefont {Grudka}},
  \bibinfo {author} {\bibfnamefont {R.}~\bibnamefont {Horodecki}}, \bibinfo
  {author} {\bibfnamefont {J.}~\bibnamefont {\L{}odyga}}, \ and\ \bibinfo
  {author} {\bibfnamefont {J.}~\bibnamefont {Wychowaniec}},\ }\href {\doibase
  10.1103/PhysRevA.99.022304} {\bibfield  {journal} {\bibinfo  {journal} {Phys.
  Rev. A}\ }\textbf {\bibinfo {volume} {99}},\ \bibinfo {pages} {022304}
  (\bibinfo {year} {2019})}\BibitemShut {NoStop}%
\bibitem [{\citenamefont {Vairogs}\ \emph {et~al.}(2022)\citenamefont
  {Vairogs}, \citenamefont {Katariya},\ and\ \citenamefont
  {Wilde}}]{Vairogs22}%
  \BibitemOpen
  \bibfield  {author} {\bibinfo {author} {\bibfnamefont {C.}~\bibnamefont
  {Vairogs}}, \bibinfo {author} {\bibfnamefont {V.}~\bibnamefont {Katariya}}, \
  and\ \bibinfo {author} {\bibfnamefont {M.~M.}\ \bibnamefont {Wilde}},\ }\href
  {\doibase 10.1103/PhysRevA.105.052434} {\bibfield  {journal} {\bibinfo
  {journal} {Phys. Rev. A}\ }\textbf {\bibinfo {volume} {105}},\ \bibinfo
  {pages} {052434} (\bibinfo {year} {2022})}\BibitemShut {NoStop}%
\end{thebibliography}%

\end{document}